\documentclass[a4paper,12pt]{scrartcl}

\usepackage[utf8]{inputenc}

\usepackage{amsmath,bm,amsthm,graphicx}
\usepackage[commonsets]{skmath}

\usepackage{graphicx,color}

\usepackage{booktabs}
\usepackage{siunitx}

\usepackage{csquotes}

\usepackage{bera}
\usepackage[charter]{mathdesign}

\usepackage{csquotes}

\usepackage{cite}
\usepackage{xspace,microtype}

\usepackage[english]{babel}

\usepackage[colorinlistoftodos,shadow]{todonotes}

\usepackage{url}

\newcommand{\adv}{{\mathcal A}}

\begin{document}

\title{A new time-stepping strategy
  and boundary treatment to improve recent 2d traffic model}
\author{Friedemann Kemm\thanks{{Brandenburg
    University of Technology (BTU) Cottbus-Senftenberg}, {Institute of Mathematics}, {Platz der Deutschen
    Einheit 1}, {03046}, {Cottbus}, {Germany}}}

\maketitle

\paragraph*{Keywords:}
\label{sec:keywords:}
Macroscopic model for urban traffic;
Partial differential equations;
Simulation and validation;
Conservation laws;
Finite volume method;
Time step restriction

\begin{abstract}
  We show how a recently published 2d model for traffic flow can be
  further improved. Besides other improvements and simplifications, we
  present not only a method to compute the necessary time step
  restrictions, but also a subcycling for the inflow and outflow. This
  drastically reduces computational cost on large domains with coarse
  grids, i.\,e.\ for simulations of a whole region instead of a small
  part of a city or town.
\end{abstract}

\section{Introduction}
\label{sec:introduction}

Since the automobile started to play a major role in traffic and
transport, traffic simulations became more and more important, in
order to plan and control traffic flow in congested areas. The
research in the past basically split into two main directions: agent
based models like the examples summarized in~\cite{NGUYEN2021100486}
and~\cite{https://doi.org/10.1155/2022/1252534}, and continuous
models. While agent based models typically rely on systems of ODEs or
even simpler models for the movement of the agents and their
interaction, continuous models employ PDEs, most notably the LWR
model, named after Lighthill and
Whitham~\cite{doi:10.1098/rspa.1955.0089} and
Richards~\cite{doi:10.1287/opre.4.1.42}.  From these basic concepts,
research on traffic flow modeling has spread out in all possible
directions~\cite{VANWAGENINGENKESSELS2015445,kessels2019traffic,app14188456}.
Although continuous models, in contrast to the microscopic agent based
models, are macroscopic, it is difficult to save computational mostly
because the LWR model is one-dimensional and, thus, has to be employed
on a single lane. While there, naturally, are extensions of the model
for multi-lane roads, they still share a prominent feature with agent
based models: the complete resolution of the street network.
%
%
When computation time is crucial, this leads to a disadvantage of the
continuous models. Quite recently some research groups tried to
overcome this issue by suggesting models that do not resolve the
complete street network. One subclass are reservoir based models like
in Mariotte et~al.~\cite{mariotte2020calibration}, another are
continuous 2d-models
like~\cite{sossoe2015traffic,mollier2019two,TUMASH2022374}. Both
subclasses allow for control of the computation time by changing the
spatial resolution of the domain of interest.
A lot of research in continuous models went into the fine tuning of
the fundamental diagram relating density and flux to each
other~\cite{greenshields1935study,NEWELL1993281,NEWELL1993289,CASTILLO1995391,doi:10.1080/18128602.2011.556680}. These
concepts were generalized for multimodal traffic by several authors~\cite{loder2019general,Huang31122022}.
Also, the LWR model itself was adapted to work with multiple transport
modes~\cite{Wierbos19022021,10422168,BENZONI-GAVAGE_COLOMBO_2003,WONG2002827}.
As mentioned above, other research was directed towards transferring
the LWR model to two space
dimensions~\cite{AGHAMOHAMMADI2019380,AGHAMOHAMMADI202099,DU201396,JIANG201841,JIANG2011343,Lin03072017,della2010distributed,TUMASH2022374,tumash2021traffic}.

In this study, we take a deeper look into the model proposed by Tumash
et~al.~\cite{TUMASH2022374,tumash2021traffic} which, although it has
already obtained about 20 citations, has not yet seen any further
development. It is mainly aimed to be used on Cartesian grids since it
considers partial densities for all four cardinal directions. Note
that, due to the small overhead, Cartesian grids are advantageous in
terms of computation time.  While the resulting system is hyperbolic,
the sum of the equations for the cardinal directions resembles a
hyperbolic balance law for the full density. The parameters for the
transport part are computed from the geometry of the street network,
the parameters for the change between cardinal directions from
measured turning ratios. The original goal of Tumash et~al.\ is to
create a tool for traffic control in a relatively small street network
like the old town of a city~\cite{tumash2021traffic,9683167}. This has
consequences for the use of the model. They aim at higher resolutions
in order to clearly resolve important features of the traffic so that
urban control mechanisms can react to
them. In~\cite[Remark~2.]{TUMASH2022374}, they require \enquote{\dots
  As a general rule of thumb, one should take cells small enough such
  that the density profile inside can be assumed homogeneous (as in
  LWR).} While for their purpose, such fine grids are a good choice,
they are disadvantageous in other contexts like traffic planning for
larger regions. If the simulation is used as a tool in a planning
council, this would lead to very high computation times that might
easily exceed the duration of one meeting. Thus, we investigate how
the model can be used on coarser grids. While in their original work,
Tumash
et~al.~\cite{TUMASH2022374,lyurlik_multidirectional_traffic_model}
employ a pre-chosen fixed time step of 0.1~seconds, we want to provide
an automated choice of the time step that guarantees numerical
stability and avoids negative densities. It should be stressed that on
fine grids the restrictions from the transport part dominate, whereas
on coarse grids the restrictions from mixing and inflow/outflow
determine the size of the time step. Since the inflow and outflow is
expected to pose the most severe restriction, and since the mixing
between the cardinal directions only poses a restriction if we require
positivity for all partial densities, we provide a subcycling strategy
for the inflow and outflow of vehicles into and out of the street
network under consideration. In addition, there are some minor changes
we propose: using a 2d-density of the state variables in
order to achieve independence of the scaling of the results from the
grid resolution, and an improved boundary treatment.

The paper is organized as follows: First we discuss the original model
in Section~\ref{sec:revis-model-tumash}. In this course, we also show
how to transfer to a 2d-density of the state
variables. Section~\ref{sec:boundary-treatment} is dedicated to the
boundary treatment and Section~\ref{sec:time-step-restr} to the new
time stepping strategy and the subcycling. Finally we discuss the
results of the different variants of the method in
Section~\ref{sec:numerical-results} before we give some conclusions
and an outlook on possible further modifications.



\section{Revisiting the model by Tumash et al.}
\label{sec:revis-model-tumash}

Note that in the description of the model, we not only use the paper
by Tumash et~al.~\cite{TUMASH2022374}, but also Tumash's
PhD-thesis~\cite{tumash2021traffic} and their original
code~\cite{lyurlik_multidirectional_traffic_model}. At points where
there is a discrepancy between these sources, or when the paper and
even the thesis are unclear in the description, we rely on the code
since it obviously is the tool used to generate the numerical results
shown in both the paper and the thesis.

\subsection{The model PDEs}
\label{sec:model-pdes}

The core of the model comprises splitting the traffic into partial
traffic densities: \textbf{N}orth bound, \textbf{E}ast bound,
\textbf{W}est bound, and \textbf{S}outh bound (\(\rho_N,\,\rho_E,\,\rho_W,\,\rho_S\)), the so-called NEWS
framework.

In the following we take, as per usual, west--east as the
\(x\)-direction and south--north as the \(y\)-direction. With these,
the model by Tumash et~al.~\cite{TUMASH2022374} reads as
\begin{equation}\label{eq:1}
  \begin{split}
    {\rho_N}_t + \left(\bar{\cos{\theta}}_N\, \Phi_N\right)_x
    + \left(\bar{\sin{\theta}}_N\, \Phi_N\right)_y
    & = \frac{1}{L} \left( {\Phi_N^{\text{in}} -
      \Phi_N^{\text{out }}}\right) +
      \frac{1}{L} \left( {\Phi_N^{\text{source}}
      - \Phi_N^{\text{sink}}}\right)\;,\\
    {\rho_E}_t + \left(\bar{\cos{\theta}}_E\, \Phi_E\right)_x
    + \left(\bar{\sin{\theta}}_E\, \Phi_E\right)_y
  & = \frac{1}{L} \left( {\Phi_E^{\text{in}} -
    \Phi_E^{\text{out }}}\right) +
    \frac{1}{L} \left( {\Phi_E^{\text{source}}
    - \Phi_E^{\text{sink}}}\right)\;,\\
    {\rho_W}_t + \left(\bar{\cos{\theta}}_W\, \Phi_W\right)_x
    + \left(\bar{\sin{\theta}}_W\, \Phi_W\right)_y
    & = \frac{1}{L} \left( {\Phi_W^{\text{in}} -
      \Phi_W^{\text{out }}}\right) +
      \frac{1}{L} \left( {\Phi_W^{\text{source}}
      - \Phi_W^{\text{sink}}}\right)\;,\\
    {\rho_S}_t + \left(\bar{\cos{\theta}}_S\, \Phi_S\right)_x
    + \left(\bar{\sin{\theta}}_S\, \Phi_S\right)_y
    & = \frac{1}{L} \left( {\Phi_S^{\text{in}} -
      \Phi_S^{\text{out }}}\right) +
      \frac{1}{L} \left( {\Phi_S^{\text{source}}
      - \Phi_S^{\text{sink}}}\right)
  \end{split}
\end{equation}
with geometry
terms~\(\cos{\theta}_N,\,\dots,\sin{\theta}_N,\,\dots,L\) (cf.\ Section~\ref{sec:geometry-terms}),
where we split the right-hand side into mixing terms
with
\begin{equation}\label{eq:19}
  \begin{split}
    \Phi_N^{\text{in}} & = \Phi_{NN} + \Phi_{EN} + \Phi_{WN} + \Phi_{SN}\;,
    \\
    \Phi_N^{\text{out}} & = \Phi_{NN} + \Phi_{NE} + \Phi_{NW} + \Phi_{NS}
  \end{split}
\end{equation}
and sources and sinks due to inflow and outflow of vehicles into and
out of the computational domain. 
The turning fluxes in equation~\eqref{eq:19} are given by
\begin{equation}\label{eq:54}
  \Phi_{EN} = \min{{\bar \alpha_{EN}} \bar D_E, {\bar \beta_{EN}} \bar S_N}\;,
\end{equation}
where~\(\bar D_E\) and~\(\bar S_N\) are the averaged demand coming
from east and the averaged supply for traffic heading north,
and~\({\bar \alpha_{EN}}\) and~\({\bar \beta_{EN}}\) are the averaged
turning ratios for traffic coming from or going to the respective
cardinal direction. For the other directions, the mixing fluxes are
computed accordingly.

Demand and supply are computed according to the cell transmission
method by Daganzo as described in~\cite{daganzo1994cell,daganzo1995cell,lebacque2002first,LECLERCQ2007701,10422168}:
\begin{align}
  \label{eq:37}
   D(\rho) & =
                 \begin{cases}
                   {\rho}\, { v}( \rho) & \text{for}\quad 0\leq  \rho
                   \leq \rho_\text{crit} \\
                   \phi_{\text{max}} & \text{for} \quad 
                   \rho_\text{crit} <  \rho \leq  \rho_\text{max}
                 \end{cases}\;, \\
   S(\rho) & =
                 \begin{cases}
                   \phi_{\text{max}} & \text{for}\quad 0\leq  \rho
                                       \leq \rho_\text{crit} \\
                   {\rho}\, { v}( \rho) & \text{for} \quad 
                   \rho_\text{crit} <  \rho \leq  \rho_\text{max}
                 \end{cases}\;,
\end{align}
where the critical density is defined
by~\(\rho_\text{crit}\, v(\rho_\text{crit}) = \phi_{\text{max}}\)
with~\(v(\rho)\) being the speed of the traffic.
These will be discussed in more detail in
Section~\ref{sec:speed-vehicles}, where a specific fundamental diagram
(FD) for the relation between density and flux is given.

Since we only have to deal with averaged quantities once we are in the
NEWS framework and especially on the computational grid, we drop the
overlines wherever possible. We only keep them for the sine and cosine
terms in the equations in order to avoid confusion. Thus, we get
\begin{equation}\label{eq:2}
  \Phi_{EN} = \min{{ \alpha_{EN}}  D_E, { \beta_{EN}}  S_N}\;. 
\end{equation}
The fluxes for the sources and sinks are given by
\begin{align}
  \vec \Phi^\text{source} (\vec \rho)
  & = \min{\vec D^\text{source},\vec S (\vec \rho)}  \;, \label{eq:io27}\\
    \vec \Phi^\text{sink} (\vec \rho)
  & = \min{\vec D (\vec \rho),\vec S^\text{sink}} \label{eq:io28} \;.
\end{align}
With the projection coefficients that we will detail in
Section~\ref{sec:geometry-terms}, demand and supply for sources and
sinks for the cardinal directions at a single intersection can be
written as
\begin{equation}
  \label{eq:46}
  D_N^\text{source} =\sum_{j=1}^{n_\text{out}}  p_{\theta_j}^N
  D_j^\text{source}\;,\qquad 
  S_N^\text{sink} = \sum_{j=1}^{n_\text{out}}  p_{\theta_j}^N
  S_j^\text{sink}\;,
\end{equation}
where the sum is taken over all streets outbound from that
intersection.

From all of this, it becomes obvious that besides the mixing terms,
there is no coupling of the equations for the partial densities
in~\eqref{eq:1}.


\subsection{The parameters}
\label{sec:how-comp-param}

We first have to compute the parameters in the NEWS framework for each
intersection~\((x_k,y_k)\) separately. Then they are interpolated at
the barycenters of the grid cells. For any point~\((x,y)\) in the
computational domain, the value of a quantity~\(Q\) would be computed
by the so-called inverse distance weighting:
\begin{equation}
  \label{eq:39}
  Q(x,y) = \frac{\sum_{k=1}^K Q(x_k,y_k)
    \exp{-\mu\sqrt{(x-x_k)^2+(y-y_k)^2}}}{\sum_{k=1}^K
    \exp{-\mu\sqrt{(x-x_k)^2+(y-y_k)^2}}}
\end{equation}
with a fixed parameter~\(\mu>0\) and the summation running over all
intersections. \enquote{Inverse} because the exponent is a negative
multiple of the distance between the point~\((x,y)\) and the location
of the k-th intersection~\((x_k,y_k)\).  Since this is the same
procedure for all parameters, in the following we only need to
consider how they are derived for a single intersection.

\subsubsection{Geometry terms}
\label{sec:geometry-terms}

The computation of the geometrical terms in~\cite{TUMASH2022374,lyurlik_multidirectional_traffic_model,tumash2021traffic} can
be simplified using linear algebra. The cosine of the angle between
two vectors/directions is
\begin{equation}
  \label{eq:31}
  \cos{
    \begin{pmatrix}
      x \\ y 
    \end{pmatrix},
    \begin{pmatrix}
      w \\ z
    \end{pmatrix}}
  = \frac{
    \begin{pmatrix}
      x \\ y 
    \end{pmatrix} \cdot
    \begin{pmatrix}
      w \\ z
    \end{pmatrix}}
  {\left|
      \begin{pmatrix}
        x \\ y 
      \end{pmatrix}\right| \cdot
    \left|
      \begin{pmatrix}
        w \\ z
      \end{pmatrix}\right|}
  = \frac{xw + yz}{\sqrt{x^2 + y^2}\cdot \sqrt{w^2 + z^2}}\;. 
\end{equation}
Thus, for flow in the direction~\(
(\xi, \eta)
\), the cosine and sine become
\begin{equation}
  \label{eq:32}
  \cos\theta = 
  \cos{
    \begin{pmatrix}
      \xi \\ \eta
    \end{pmatrix},
    \begin{pmatrix}
      1 \\ 0
    \end{pmatrix}}
  = \frac{\xi}{\sqrt{\xi^2 + \eta^2}}\;,\quad
  \sin\theta =
  \sin{
    \begin{pmatrix}
      \xi \\ \eta
    \end{pmatrix},
    \begin{pmatrix}
      1 \\ 0
    \end{pmatrix}}
  =  \frac{\eta}{\sqrt{\xi^2 + \eta^2}}\;,
\end{equation}
Where~\(\theta\) is the angle between~\((\xi,\eta)\) and the
\(x\)-axis, which resembles here the east direction.  This allows us
to simplify the projection coefficients
in~\cite[equation~(10)]{TUMASH2022374} as
\begin{equation}
  \label{eq:34}
  p_{(\xi, \eta)}^N = \frac{\max{\eta,0}}{\abs{\xi} + \abs{\eta}}\;,\quad
  p_{(\xi, \eta)}^E = \frac{\max{\xi,0}}{\abs{\xi} + \abs{\eta}}\;,\quad
  p_{(\xi, \eta)}^W = \frac{\min{\xi,0}}{\abs{\xi} + \abs{\eta}}\;,\quad
  p_{(\xi, \eta)}^S = \frac{\min{\eta,0}}{\abs{\xi} + \abs{\eta}}\;.
\end{equation}
Since the street network is given in Cartesian coordinates, this saves
computation time and reduces rounding errors.

These coefficients are also used in the computation of the other
parameters in the NEWS framework, especially the averaged cosine and
sine values in system~\eqref{eq:1}. E.\,g.\ for the north direction,
we have to compute
\begin{equation}
  \label{eq:36}
  \bar{\cos{\theta}}_N = \frac{\sum_{j=1}^{n_\text{out}} p_{\theta_j}^N
    \cos{\theta_j}
    \phi_{\text{max},j}}{\sum_{j=1}^{n_\text{out}}   p_{\theta_j}^N\phi_{\text{max},j}}\;,\qquad 
  \bar{\sin{\theta}}_N =\frac{\sum_{j=1}^{n_\text{out}} p_{\theta_j}^N
    \sin{\theta_j}
    \phi_{\text{max},j}}{\sum_{j=1}^{n_\text{out}}  p_{\theta_j}^N\phi_{\text{max},j}}\;,
\end{equation}
where \(j=1,\dots,n_\text{out}\) are the indices of the streets
exiting from the intersection under consideration. Thus, we get
averaged sines and cosines at every intersection. While here we have
omitted the indices of the intersections themselves, we will later
need them in the computation of the values on the computational grid.

However, for that purpose, we need to turn the values in the cells
into values at the cell faces. Instead of computing them at the
midpoints of the cell faces, we follow Tumash
et~al.~\cite{lyurlik_multidirectional_traffic_model,TUMASH2022374} and
first get the values in the barycenters of the cells. Afterwards, we
compute the arithmetic mean of the values in the cells left and right
of the cell face for the cosine and below and above the cell face for
the sine.  This yields
\begin{align}
    \label{eq:33}
      \bar{\cos{\theta}}_{N}[i+\tfrac 1 2,j],&&
      \bar{\cos{\theta}}_{E}[i+\tfrac 1 2,j],&&
      \bar{\cos{\theta}}_{W}[i+\tfrac 1 2,j],&&
      \bar{\cos{\theta}}_{S}[i+\tfrac 1 2,j]&& \forall\ i,j\;,\\
    \label{eq:35}
      \bar{\sin{\theta}}_{N}[i,j+\tfrac 1 2],&&
      \bar{\sin{\theta}}_{E}[i,j+\tfrac 1 2],&&
      \bar{\sin{\theta}}_{w}[i,j+\tfrac 1 2],&&
      \bar{\sin{\theta}}_{S}[i,j+\tfrac 1 2]&& \forall\ i,j\;.
\end{align}
These are used later when computing the advective terms.

The length parameter~\(L\) for an intersection is computed via the
lengths~\(l_j\) of the outbound streets and the maximal
density~\(\rho_{\text{max},j}\), which in turn depends on the number
of available lanes:
\begin{equation}
  \label{eq:38}
  L = \frac{\sum_{j=1}^{n_\text{out}} \rho_{\text{max},j}\,
    l_j}{\sum_{j=1}^{n_\text{out}} \rho_{\text{max},j}}\;.
\end{equation}


\subsubsection{Turning ratios}
\label{sec:turning-ratios}

We start our computations with the measured turning ratios for each
intersection, i.\,e.\ numbers~\(\alpha_{i,j}^{(k)}\in[0,1]\) that are
nonzero if it is possible to turn from street~\(i\) to street~\(j\) at
intersection~\(k\), and if there are actually drivers who do turn in
this direction. Evidently, we have
\begin{equation}
  \label{eq:22}
  \sum_j \alpha_{i,j}^{(k)} = 1\qquad \forall\quad i,k\;. 
\end{equation}
While these indicate the demand for turning at the intersections, we
also need a measure for the supply. This is done with the supply
ratios~\(\beta_{i,j}^{(k)}\in[0,1]\), which are computed as
\begin{equation}
  \label{eq:29}
  \beta_{i,j}^{(k)} = \frac{\alpha_{i,j}^{(k)}
    \Phi_{\text{max},i}^\text{in}}{\sum_l \alpha_{l,j}^{(k)}
    \Phi_{\text{max},l}^\text{in}}\;. 
\end{equation}
Obviously, similar to equation~\eqref{eq:22}, we have
\begin{equation}
  \label{eq:30}
  \sum_i \beta_{i,j}^{(k)} = 1\qquad \forall\quad i,k\;. 
\end{equation}
Next, we can compute the turning ratios with respect to the cardinal
directions by using the projection coefficients as defined in
equation~\eqref{eq:34} for the according streets. Here, we show an
example of traffic turning from eastbound to northbound:
\begin{align}
  \label{eq:44}
  \alpha_{EN} & = \frac{ \sum_{j=1}^{n_\text{out}}  p_{\theta_j}^N
                \sum_{i=1}^{n_\text{in}} \alpha_{i,j}  p_{\theta_i}^E
                \Phi_{\text{max},\,i}}{\sum_{i=1}^{n_\text{in}}
                p_{\theta_i}^E\Phi_{\text{max},\,i}}\;, \\
  \label{eq:45}
  \beta_{EN} & = \frac{ \sum_{i=1}^{n_\text{in}}  p_{\theta_i}^E
                \sum_{j=1}^{n_\text{out}} \beta_{i,j}  p_{\theta_j}^N
                \Phi_{\text{max},\,j}}{\sum_{j=1}^{n_\text{out}}
                p_{\theta_j}^N \Phi_{\text{max},\,j}}\;,
\end{align}
where we dropped the superscripts \enquote{(in)} and \enquote{(out)}
assuming that the index~\(i\) runs only over the streets incoming to
the intersection and~\(j\) only over the streets outgoing from it. If
incoming and outgoing streets at an intersection are stored in
different variables or, better, different data structures in a code,
there won't be a problem with equal indices.


\subsubsection{Maximal speed, density, etc.}
\label{sec:maxim-speed-dens}

In general, the density of traffic heading north at an intersection is
computed via
\begin{equation}
  \label{eq:40}
  \rho_N = \sum_{i=1}^{n_\text{in}}  p_{\theta_i}^N \rho_{i}
  + \sum_{j=1}^{n_\text{out}}  p_{\theta_j}^N \rho_{j}\;,
\end{equation}
where the indices~\(i=1,\dots,n_\text{in}\)
and~\(j=1,\dots,n_\text{out}\) run over the streets incoming to and
outgoing from the intersection, respectively. This also holds true for
the maximal density~\(\rho_{\text{max},\,N}\):
\begin{equation}
  \label{eq:41}
  \rho_{\text{max},\,N} = \sum_{i=1}^{n_\text{in}}  p_{\theta_i}^N \rho_{\text{max},i}
  + \sum_{j=1}^{n_\text{out}}  p_{\theta_j}^N \rho_{\text{max},\,j}\;.
\end{equation}
The critical density~\(\rho_\text{crit}\), i.\,e.~the density below
which we expect freely flowing traffic at maximal
speed~\(v_\text{max}\), is treated likewise.
Tumash~\cite{tumash2021traffic} gives general formulas for the
maximal speed~\(v_\text{max}\) and the kinematic wave speed~\(c_K\)
in the cardinal directions at the example of northbound traffic as
\begin{align}
  \label{eq:42}
  v_{\text{max},\,N} & = \frac{
                       \sum_{i=1}^{n_\text{in}}  p_{\theta_i}^N v_i
                       \rho_{\text{crit},\,i}  +
                       \sum_{j=1}^{n_\text{out}}  p_{\theta_j}^N  v_j
                       \rho_{\text{crit},\,j} }
                       {\rho_{\text{crit},\,N}} \\
  \intertext{and}
  \label{eq:43}
  c_{K,\,N} & = \frac{
              \sum_{i=1}^{n_\text{in}}  p_{\theta_i}^N c_{K,\,i}
              (\rho_{\text{max},\,i} - \rho_{\text{crit},\,i})
              + \sum_{j=1}^{n_\text{out}}  p_{\theta_j}^N c_{K,\,j}
              (\rho_{\text{max},\,j} - \rho_{\text{crit},\,j})}
              { \rho_{\text{max},\,N} - \rho_{\text{crit},\,N}}\;,
\end{align}
But later Tumash~\cite{tumash2021traffic} drops equation~\eqref{eq:43}
in favor of a simpler approach based on the so called bilinear
fundamental diagram as described in the next section.  Note, however,
that relying on the critical density~\(\rho_{\text{crit},\,N}\)
restricts the applicability of equation~\eqref{eq:43} to a rather
small number of according fundamental diagrams anyway. It should be
stressed that Many fundamental diagrams have no such critical density,
at least if we exclude zero density.  But as already mentioned, we
don't need the critical density as well as equation~\eqref{eq:43} if
we follow Tumash et~al.~\cite{TUMASH2022374} by choosing the so-called
bilinear fundamental diagram. Then, we only need the maximal velocity
and maximal density in the NEWS framework. Normally, we would also
need either the critical density or the kinematic wave speed as a
parameter. But following Tumash et~al., we can rely solely on a single
parameter relating critical to maximal density as described in the
following Section.

\subsubsection{Speed of the vehicles}
\label{sec:speed-vehicles}

Following Tumash et~al.~\cite{TUMASH2022374}, we employ the so-called
bilinear fundamental diagram for the flux:
\begin{equation}
  \label{eq:5}
  \phi(\rho) =
  \begin{cases}
    v_\text{max}\, \rho & \text{for}\quad \rho \leq \rho_\text{crit} \\
    \abs{c_K}\,(\rho_\text{max} - \rho) & \text{otherwise}
  \end{cases}\;,
\end{equation}
where~\(c_K\) is the kinematic wave speed (celerity) at jam
density~\(\rho_\text{max}\). If~\(u=u(\rho)\) represents the actual
speed of the traffic, then
\begin{equation}
  \label{eq:6}
  c_K = \rho_\text{max}\,\td{u}{\rho}\left(
    \rho_\text{max}\right)\qquad \bigl(\ <0\ \bigr)\;. 
\end{equation}
Since the flux is continuous as a function of the density, the
critical density and the celerity at jam density are connected via
\begin{equation}
  \label{eq:7}
  \rho_\text{crit} = \frac{-c_K \rho_\text{max}}{v_\text{max} - c_K}\;. 
\end{equation}
As a consequence, by fixing~\(c_K\) we also fix~\( \rho_\text{crit}\)
and vice versa. Although, in general,~\(c_K\) is modeled via the
reaction time of drivers, Tumash et~al.\ set 
\begin{equation}
  \label{eq:8}
  \rho_\text{crit} = \gamma \rho_\text{max}
\end{equation}
with a constant~\(\gamma\) which they choose as~\(\gamma = 1/3\) for
their computations. From~\eqref{eq:7}, we compute
\begin{equation}
  \label{eq:9}
  c_K = -\,\frac{v_\text{max} \rho_\text{crit}}{\rho_\text{max} -
    \rho_\text{crit}}\;, 
\end{equation}
which with the setting in~\eqref{eq:8} becomes
\begin{equation}
  \label{eq:10}
  c_K = \frac{\gamma}{\gamma - 1}\, v_\text{max} \qquad
  \Rightarrow\qquad \abs{c_K} = \frac{\gamma}{1-\gamma}\,
  v_\text{max}\;. 
\end{equation}
In the case of~\(\gamma = 1/3\), this is
\begin{equation}
  \label{eq:11}
  \abs{c_K} = \frac{1}{2}\, v_\text{max} \;. 
\end{equation}
Although this seems very high on highways with a high speed limit and
rather low on some streets with a very strict speed limit, we think
this is sufficiently precise considering the rather large modeling
errors we accept in order to arrive at a continuous 2d model.

The maximal flow therefore reads as
\begin{equation}
  \label{eq:12}
  \phi(\rho_\text{crit}) = \gamma\, \rho_\text{max}\, v_\text{max} \;, 
\end{equation}
and demand and supply can be written as
\begin{align}
  \label{eq:3}
   D_N(\rho_N) & = \min{{ v}_N {\rho}_N,
                   \phi_{\text{max},N}}
                   = \min{{ v}_N {\rho}_N,
                   v_{\text{max},N} \rho_{c,N}}\;,  \\
  \label{eq:4}
   S_N(\rho_N) & = \min{c_K\,(\rho_{\text{max},N} -\rho_N),\,
                   \phi_{\text{max},N}}
                   = \min{c_K\,(\rho_{\text{max},N} -\rho_N),\,
                   v_{\text{max},N}\, \rho_{c,N}} 
\end{align}
for the northbound traffic. For the other cardinal directions, the
formulas are likewise. Therefore, from now on, we drop the subscript
for the cardinal direction. For further use, we rewrite these by using
a piecewise definition as
\begin{align}\label{eq:io18}
  D(\rho) & =
            \begin{cases}
              0 & \text{for}\ \rho < 0 \\
              v_\text{max}\, \rho & \text{for}\ 0\leq \rho \leq
                                  \rho_\text{crit} \\
              v_\text{max}\, \rho_\text{crit} & \text{otherwise}
            \end{cases}\;,\\[\medskipamount]
  S(\rho) & =
            \begin{cases}
              v_\text{max}\, \rho_\text{crit} & \text{for}\ 0\leq \rho \leq
                                              \rho_\text{crit} \\
              v_\text{max}\, \frac{\rho_\text{crit}(\rho_\text{max} -
              \rho)} {\rho_\text{max} - \rho_\text{crit}}
                                            &
                                              \text{for}\
                                              \rho_\text{crit} < \rho
                                              \leq  \rho_\text{max} \\
              0  & \text{otherwise}
            \end{cases}\;. \label{eq:io19}
\end{align}
Accordingly, the derivatives with respect to~\(\rho\) are
\begin{align}\label{eq:io20}
  D_\rho(\rho) & =
                 \begin{cases}
                   0 & \text{for}\ \rho < 0 \\
                   v_\text{max} & \text{for}\ 0\leq \rho \leq
                                  \rho_\text{crit} \\
                   0  & \text{otherwise}
                 \end{cases}\;,\\[\medskipamount]
  S_\rho(\rho) & =
                 \begin{cases}
                   0 & \text{for}\ 0\leq \rho \leq
                       \rho_\text{crit} \\
                   \frac{-v_\text{max}\,\rho_\text{crit}}{\rho_\text{max}
                   - \rho_\text{crit}} &
                                              \text{for}\
                                              \rho_\text{crit} < \rho
                                         \leq  \rho_\text{max} \\
                   0 & \text{otherwise}
                 \end{cases}\;, \label{eq:io21}
\end{align}

\subsubsection{Initial and boundary values}
\label{sec:init-bound-valu}

Although initial and boundary values are no actual parameters for the model, they need to be treated similarly: computed in the NEWS
framework for each intersection and finally interpolated in the
barycenters of the computational grid. 

Since we start at midnight with an empty street network, we don't need
to go into details for the initial conditions. We implemented an
approach similar to the one in~\cite{TUMASH2022374} in order to
locate the streets on a fine grid, which is then used as a means to
plot the street network over the numerical results.


The data for traffic incoming to or outgoing from the street network
are provided from measurements. Here, we again rely on the data given
by Tumash et~al.~\cite{lyurlik_multidirectional_traffic_model}. These
are then used for sources and sinks at the positions of the according
intersections. However, while in the paper they claim that no further
information on the boundaries is needed, we follow their code by
setting homogeneous Dirichlet conditions at the boundary. The
necessity for these is discussed below in
Section~\ref{sec:boundary-treatment}.

\subsection{Scaling  of the model}
\label{sec:making-model-scal}

\begin{figure}
  \centering
  \includegraphics[width=.21\linewidth]{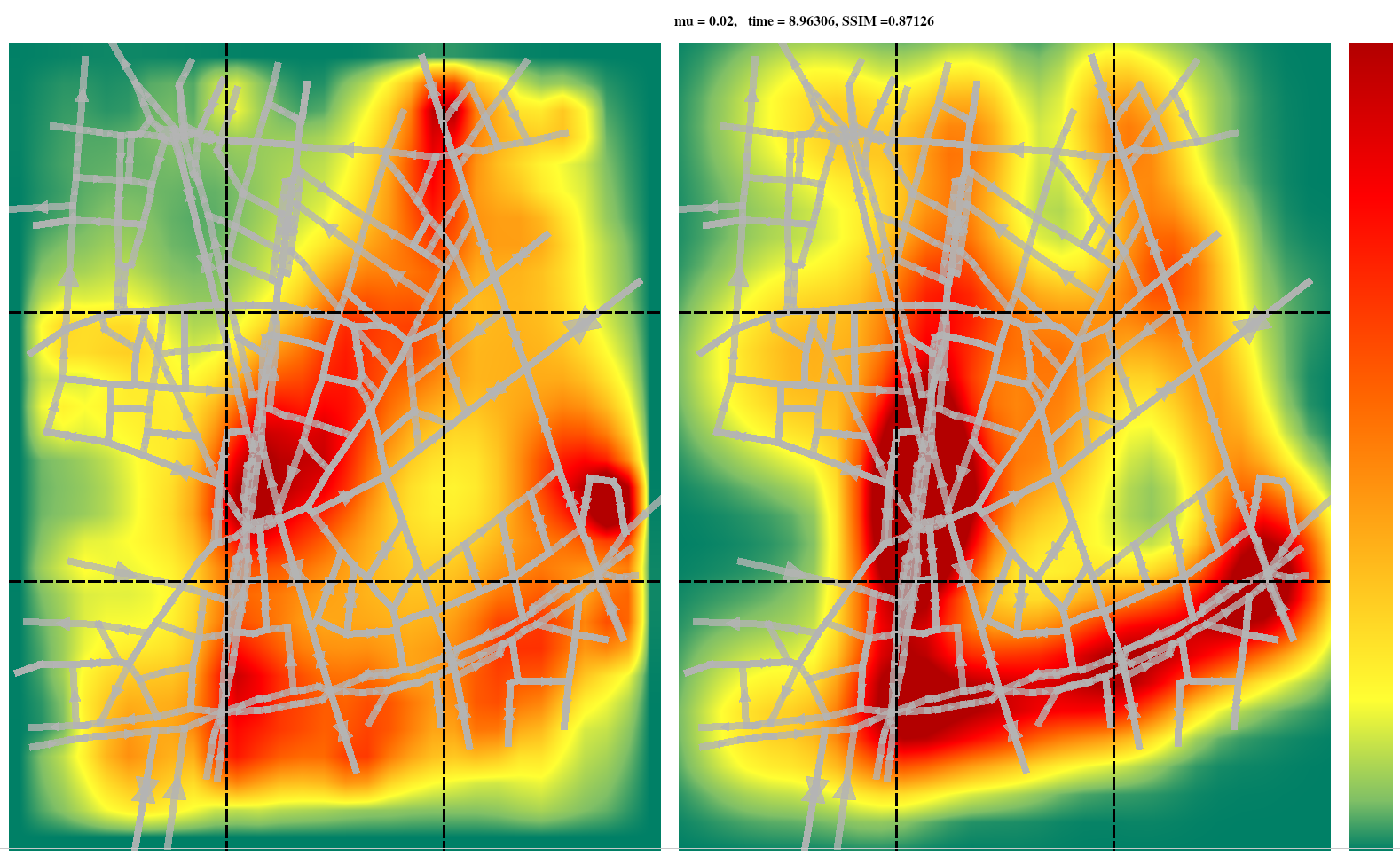}\hfill
  \includegraphics[width=.21\linewidth]{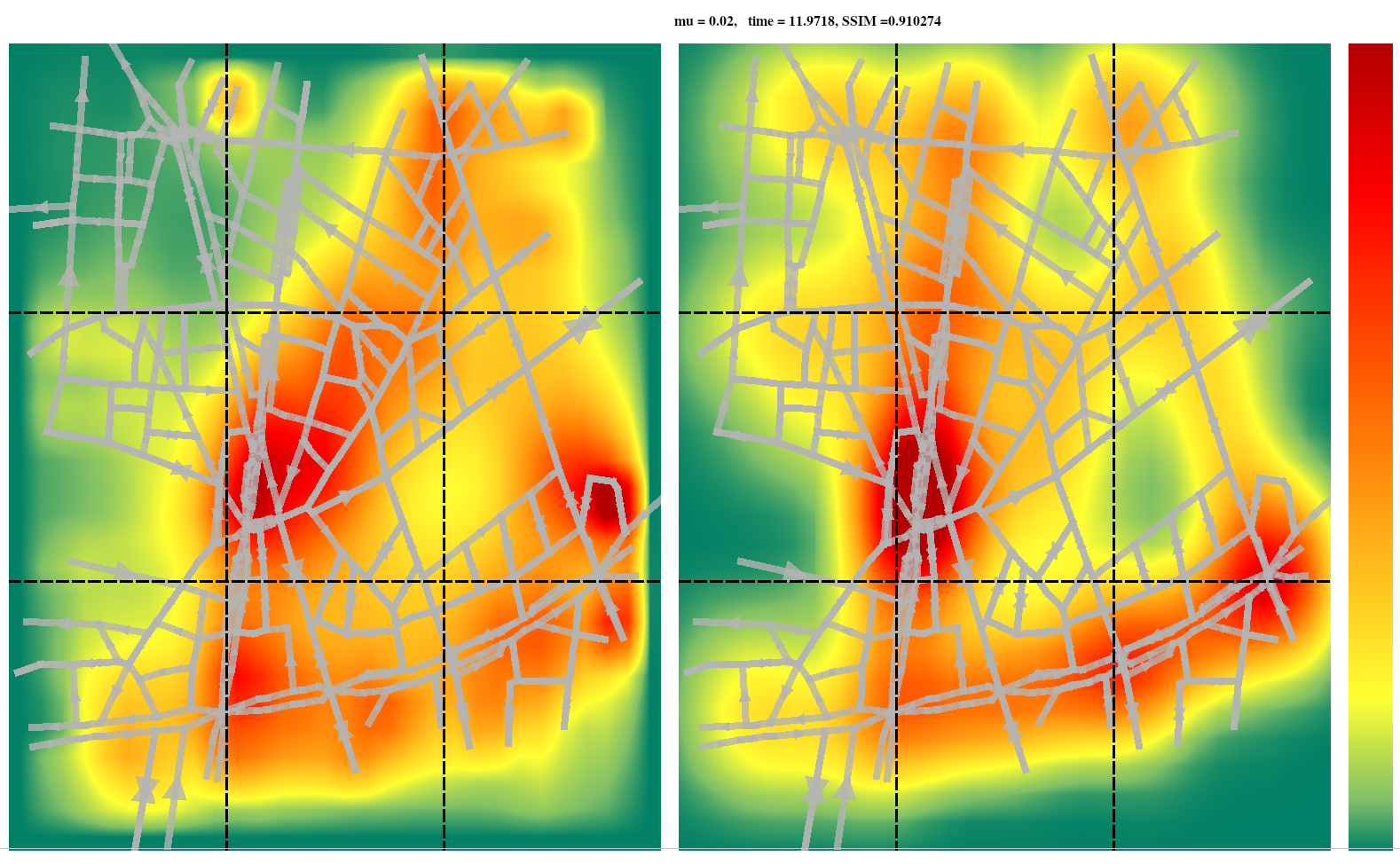}\hfill
  \includegraphics[width=.21\linewidth]{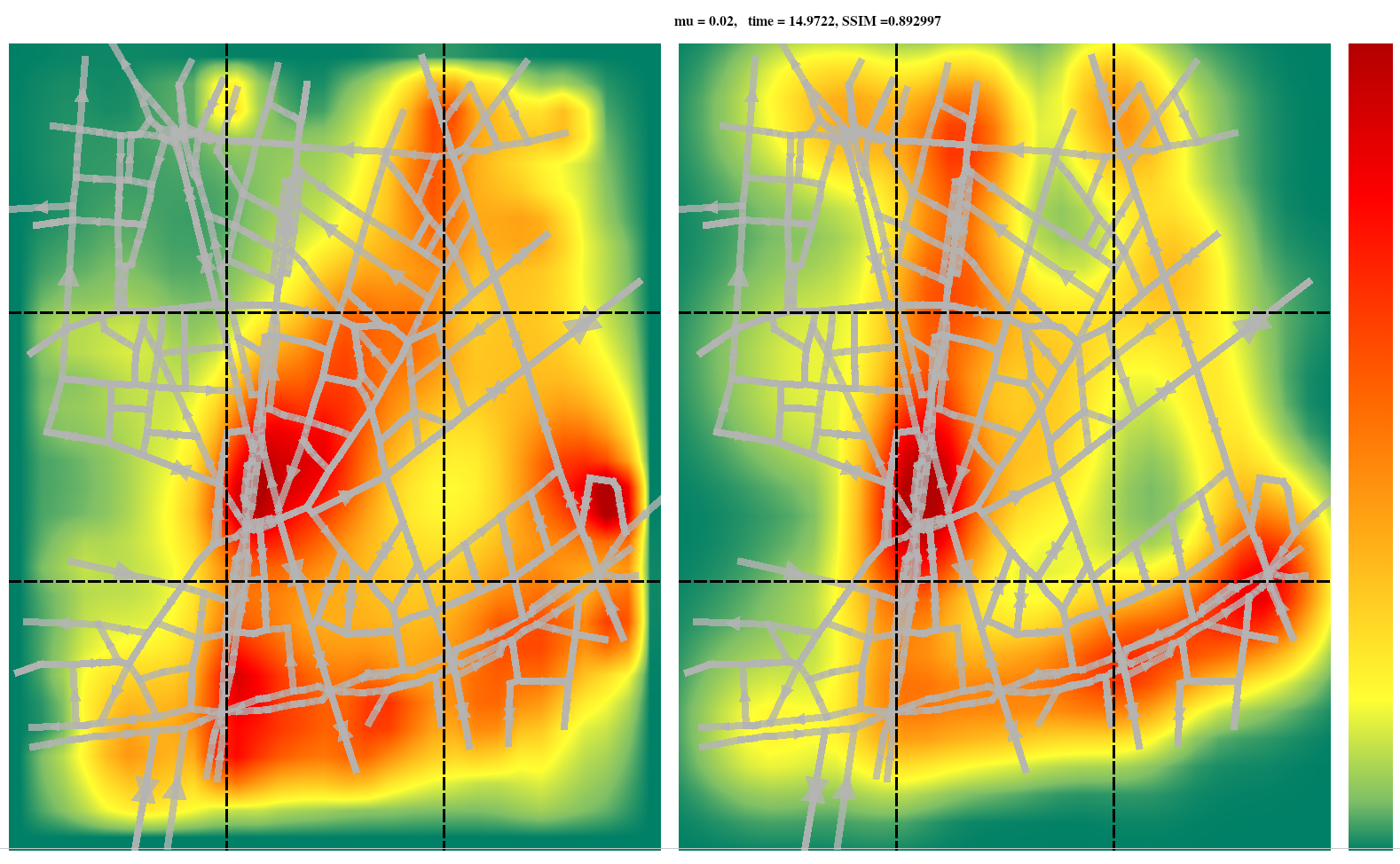}\hfill
  \includegraphics[width=.21\linewidth]{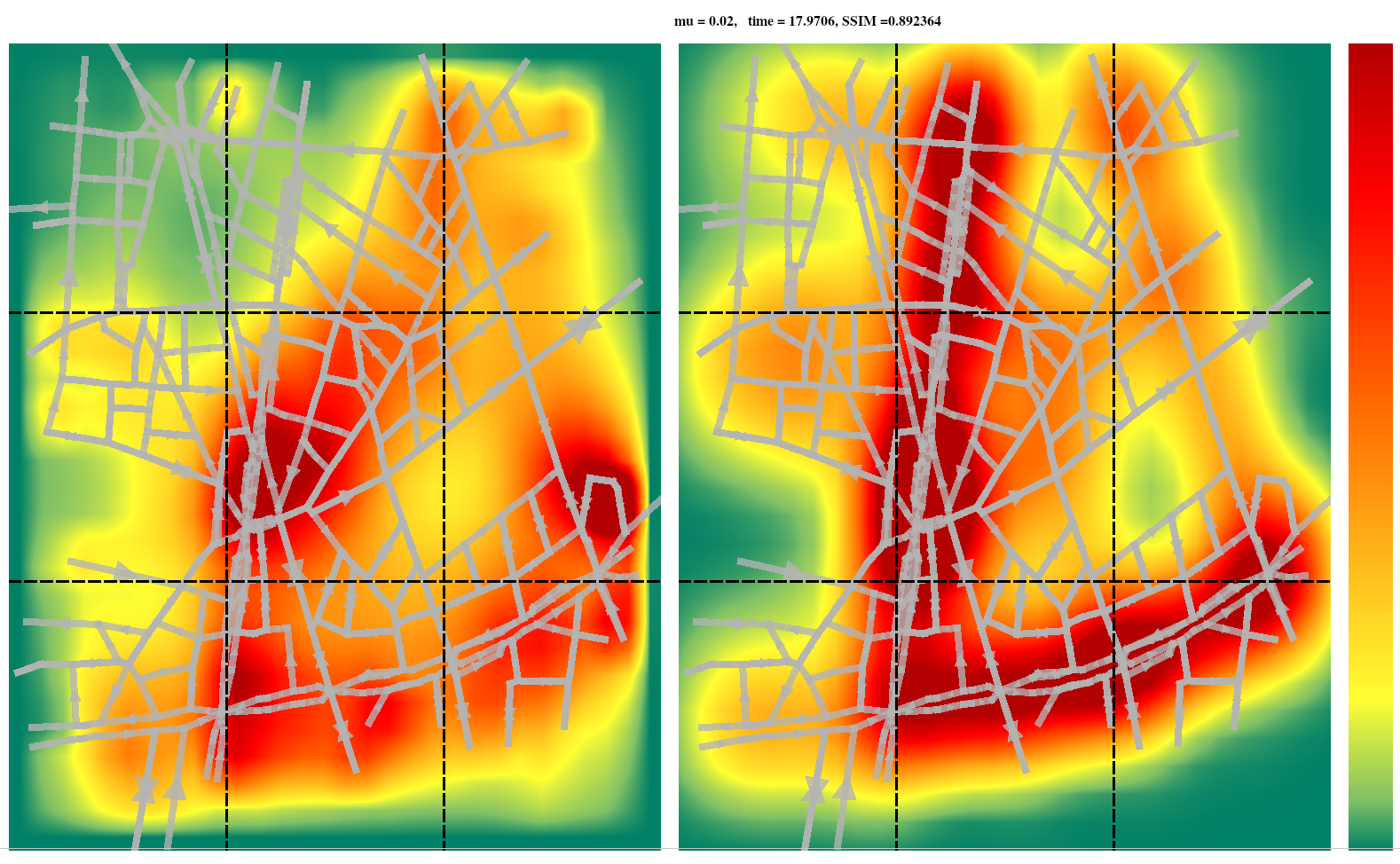}\\
  \includegraphics[width=.21\linewidth]{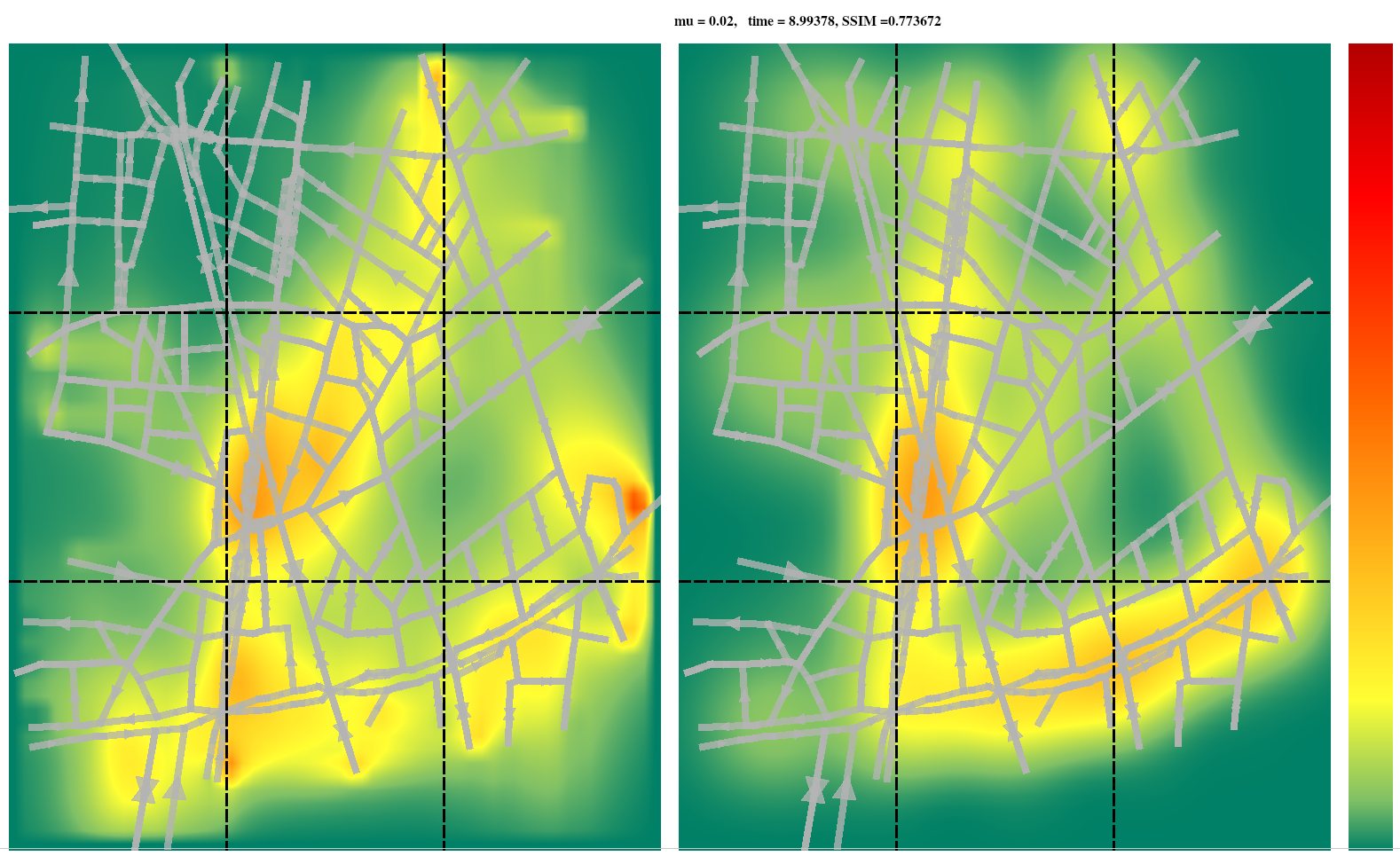}\hfill
  \includegraphics[width=.21\linewidth]{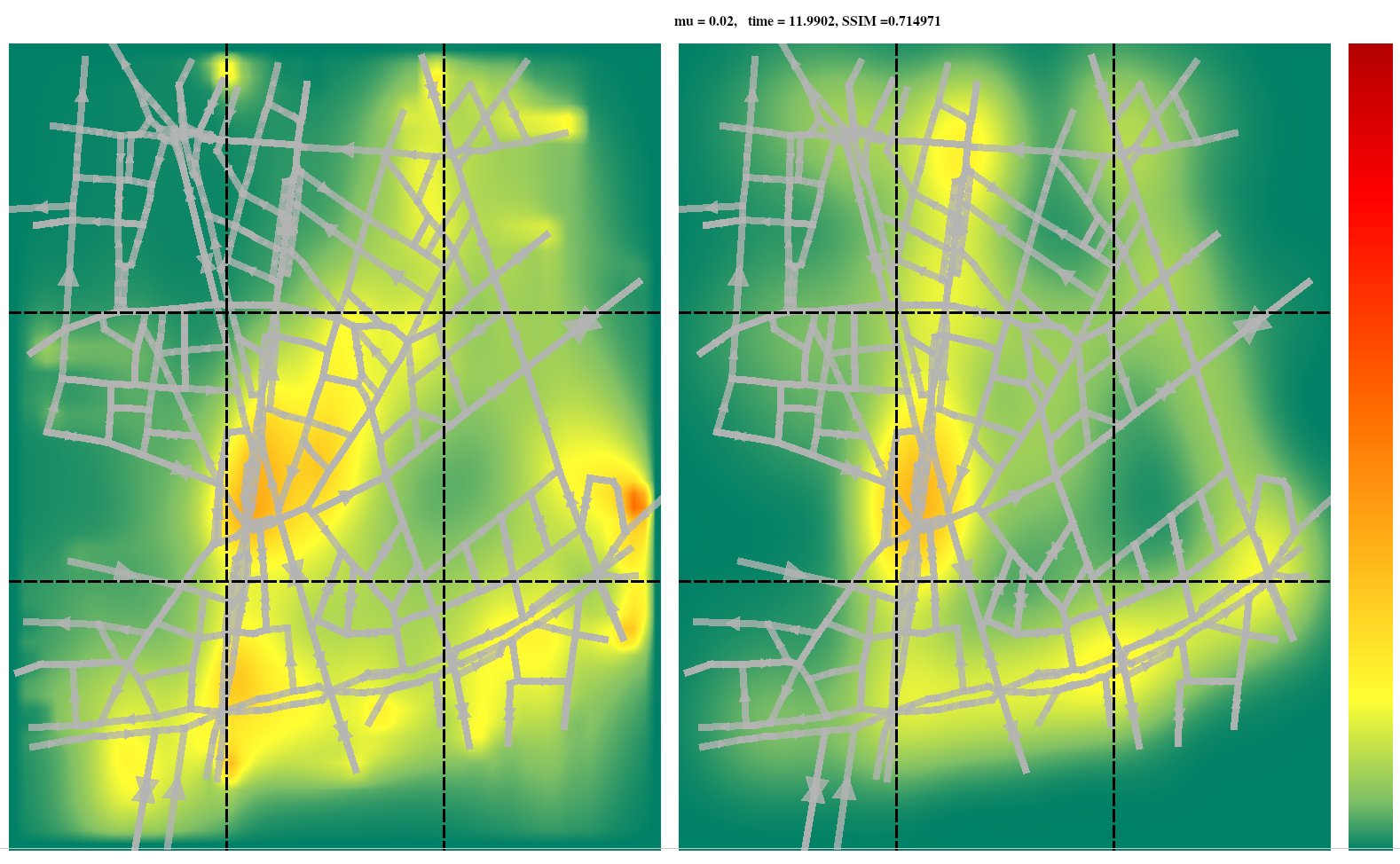}\hfill
  \includegraphics[width=.21\linewidth]{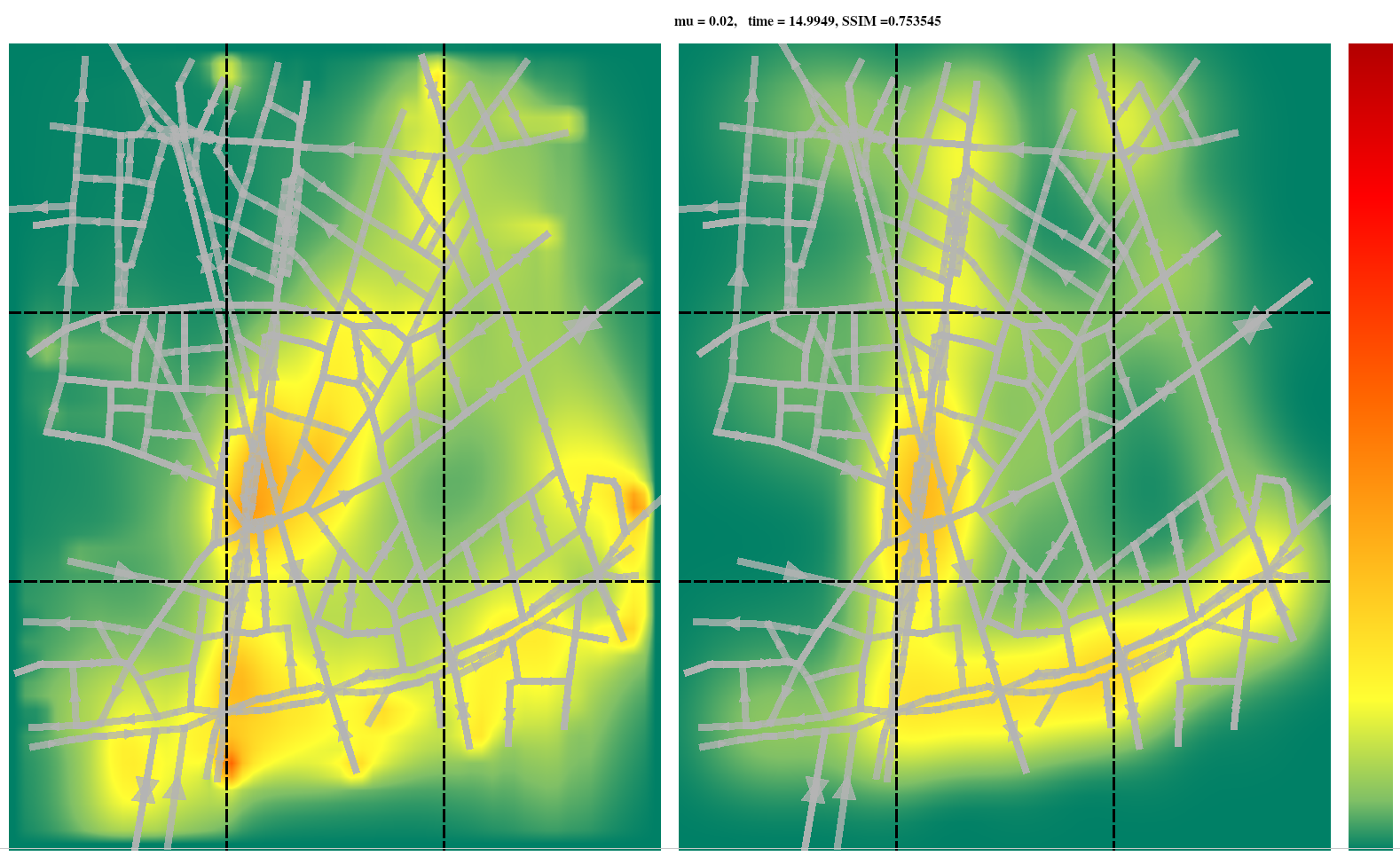}\hfill
  \includegraphics[width=.21\linewidth]{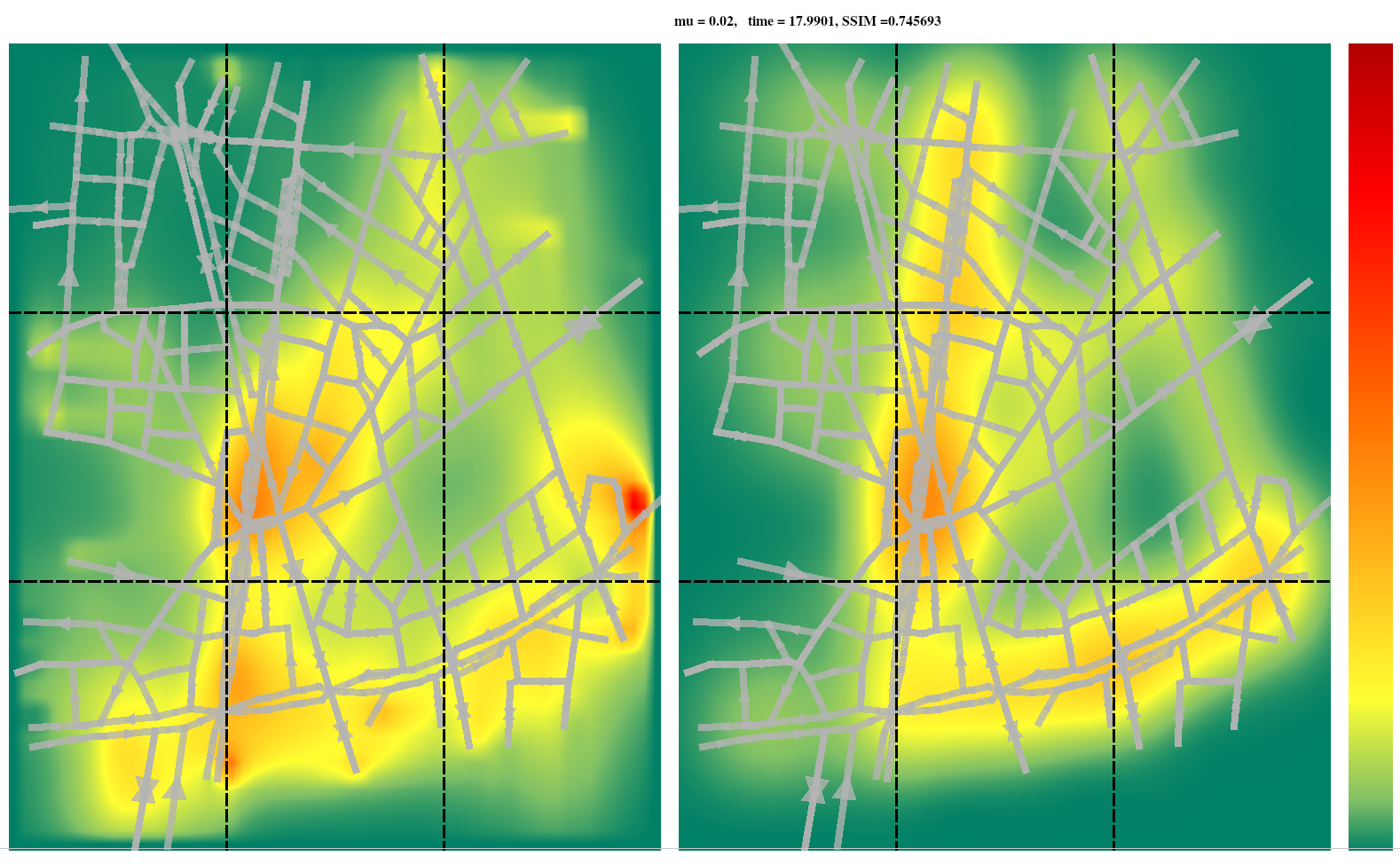}
  \caption[Results with original code]{Numerical solution (left frame)
    and measured data (right frame) with original code by Tumash et
    al.\ with 30\(\times\)30 grid cells (top row) and with 60\(\times\)60 grid cells
    (bottom row) at ca.\ 9\,am, 12\,pm, 3\,pm, and 6\,pm.}
  \label{fig:original-code}
\end{figure}

A peculiarity of the original scheme by Tumash
et~al.~\cite{TUMASH2022374} is that the scaling of the results changes
with the resolution of the numerical scheme as can be seen from
Figure~\ref{fig:original-code}. Running their code on different grid
resolutions leads to different magnitudes of traffic density in the
numerical simulation as well as in the presentation of the empirical
results. A closer look shows that they don't use a 2d density of a
\enquote{physical} quantity, but the quantity itself, in this case the
one-dimensional traffic density on the street network. We prefer the
usual approach, like in fluid dynamics, where instead of the mass,
momentum, etc. densities are considered like mass density, density of
momentum, etc. In the case of the NEWS-model, this implies that we
have to consider a 2d-density of the traffic density on the street
network. At first glance, a density of a density seems confusing,
but helps us when comparing  results on different grids.

\begin{figure}
  \centering
  \includegraphics[width=.33\linewidth]{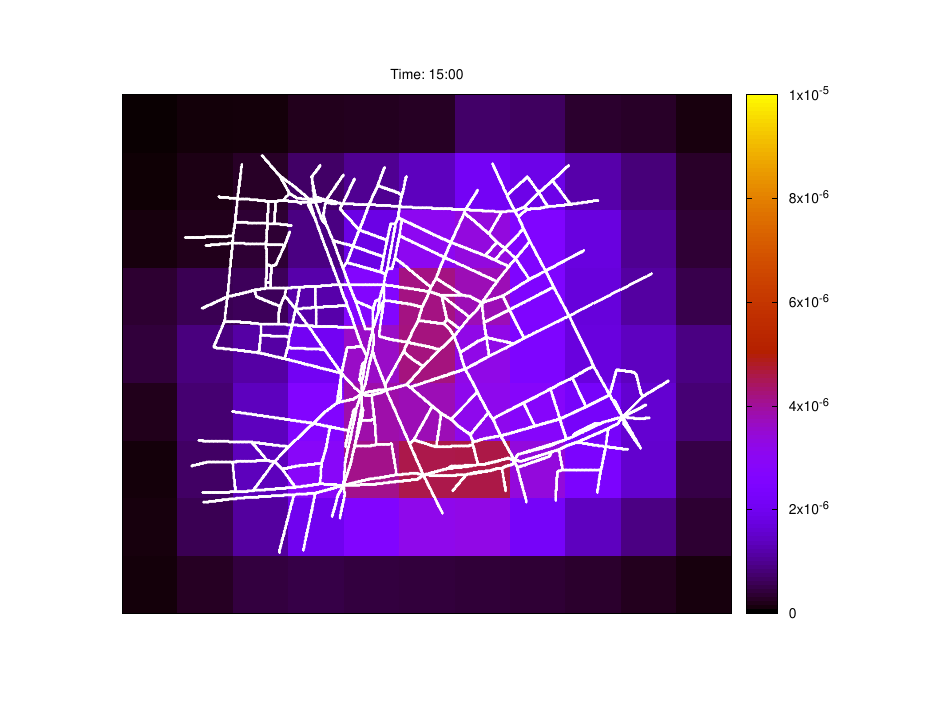}\hfill
  \includegraphics[width=.33\linewidth]{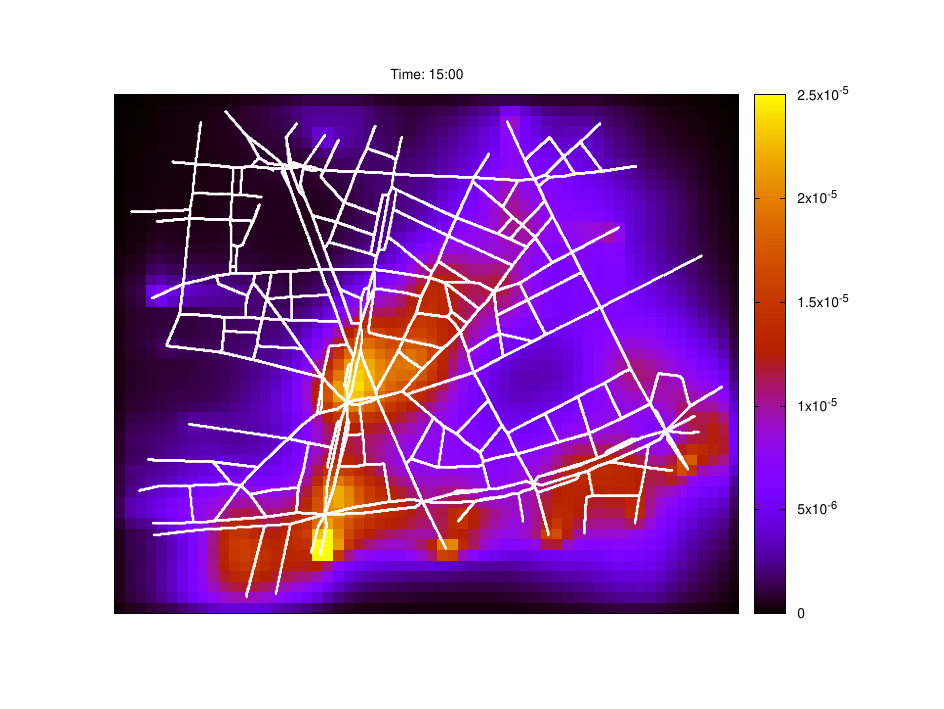}\hfill
  \includegraphics[width=
  .33\linewidth]{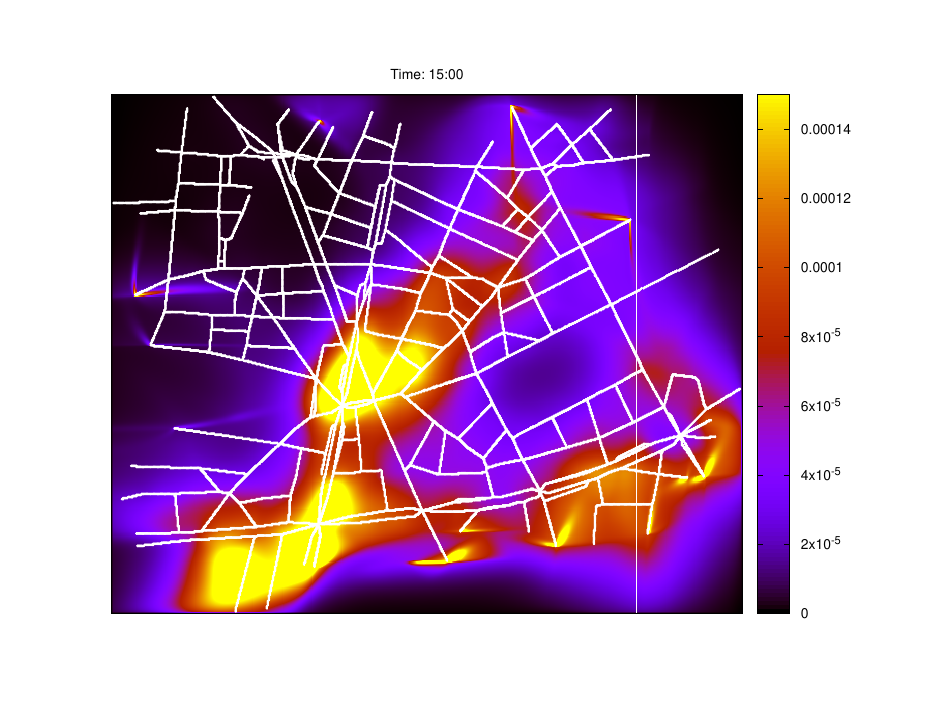}
  \caption{Results with new code for 3\,pm with (from left to right)
    12\(\times\)10, 61\(\times\)50, and 610\(\times\)500 grid cells.}
  \label{fig:resultate}
\end{figure}

It should be taken into account, however, that for very high
resolutions, we have cells without any streets and cells that are
dominated by streets. Thus, the difference between high density areas
and low density areas is much higher than on coarse grids, an effect
also visible in Figure~\ref{fig:original-code}. We will take this into
account when presenting our numerical results in
Section~\ref{sec:numerical-results} by the color range for coarse grid
solutions. An example for this is shown in Figure~\ref{fig:resultate}. 

Note that the equations~\eqref{eq:1} look essentially the same. The
main difference is that densities are now 2d-densities of traffic
densities. The most important consequence is therefore that the
sources and sinks as well as the initial data need to be treated
differently. After summing up the initial traffic in a cell and the
sources and sinks in a cell for a certain time, we then have to divide
them by the cell-size.

\subsection{Basic numerical treatment}
\label{sec:numerical-treatment}

For numerical treatment, we basically follow Tumash
et~al.~\cite{TUMASH2022374} by also employing a finite volume
method. For the sake of simplicity, Donor Cell Upwind (DCU) as
described in~\cite{leveque} is chosen to deal with the
multidimensional nature of the problem.

Without the subcycling described below in
Section~\ref{sec:subcycl-infl-outfl} computations are done in the
following way:
\begin{itemize}
\item First, in every grid cell, demand and supply have to be computed
  for every cardinal direction. This gives us values
  \begin{align}
    \label{eq:47}
    D_N[i,j], && D_E[i,j],  && D_W[i,j],  &&  D_s[i,j] && \forall\ i,j\;, \\
    S_N[i,j],   && S_E[i,j],  && S_W[i,j],  && S_S[i,j] && \forall\
                                               i,j\;. \label{eq:48}
  \end{align}
\item Next, we have to compute the leftgoing and rightgoing fluxes at
  the cell faces. For the~\(x\)-direction, this is
  \begin{align}
    \label{eq:49}
    \begin{split}
      \Phi_{N,\,R}[i+\frac 1 2,j] & = \min{D_N[i,j],S_N[i+1,j]}, \\
      \Phi_{N,\,L}[i+\frac 1 2,j] & = \min{D_N[i+1,j],S_N[i,j]}, \\
      \Phi_{N,\,R}[i,j+\frac 1 2] & = \min{D_N[i,j],S_N[i,j+1]}, \\
      \Phi_{N,\,L}[i,j+\frac 1 2] & = \min{D_N[i,j+1],S_N[i,j]}
    \end{split}
  \end{align}
  for northbound traffic. For the other cardinal directions, the
  fluxes are computed likewise. 
\item Now, we can use these and the cosine and sine values at the
  interfaces~\eqref{eq:36} to compute the advection terms at the
  interfaces:
  \begin{equation}
    \label{eq:51}
    \begin{split}
      \adv_N[i+\frac 1 2,j] & = \bar{\cos{\theta}}_{N}^+\,[i+\tfrac 1 2,j]\
                               \Phi_{N,\,R}[i+\frac 1 2,j]\ +\
                              \bar{\cos{\theta}}_{N}^-\,[i+\tfrac 1 
                              2,j]\  \Phi_{N,\,L}[i+\frac 1 2,j]\;, \\
      \adv_N[i,j+\frac 1 2] & = \bar{\sin{\theta}}_{N}^+\,[i,j+\tfrac 1
                              2]\,  \Phi_{N,\,R}[i,j+\frac 1 2]\
                              +\  \bar{\sin{\theta}}_{N}^-\,[i,j+\tfrac 1 2]\,
                               \Phi_{N,\,L}[i,j+\frac 1 2]
    \end{split}
  \end{equation}
  with the definition
  \begin{equation}
    \label{eq:52}
    q^+ = \frac 1 2 (q + \abs q) = \max{q,0}\;,\qquad q^- = \frac 1 2 (q
    - \abs q) = \min{q,0}
  \end{equation}
  for any quantity~\(q\). Thus, equation~\eqref{eq:51} gives us
  upwinding for the discretization of the transport terms.

  From these, we compute the advective update as
  \begin{equation}
    \label{eq:59}
    \Phi_{N,\text{adv}}[i,j] = - \frac{1}{\Delta x}
    (\adv_N[i+\tfrac 1 2,j] - \adv_N[i-\tfrac 1 2,j]) -
    \frac{1}{\Delta y}( \adv_N[i,j+ \tfrac 1 2] - \adv_N[i,j-\tfrac 1 2])\;,
  \end{equation}
  where~\(\Delta x\) and~\(\Delta y\) denote the grid spacing
  in~\(x\)- and~\(y\)-direction respectively. The updates for the
  other cardinal directions are computed similarly. 
\item We can also compute the mixing fluxes. For the northbound
  density, i.\,e.\ for the first equation in system~\eqref{eq:1}, this
  is done according to equations~\eqref{eq:19} and~\eqref{eq:54}, or
  more precisely equation~\eqref{eq:2} after dropping the overlines,
  which leaves us with, e.\,g.
  \begin{equation}
    \label{eq:53}
    \Phi_{EN}[i,j] = \min{{ \alpha_{EN}}  D_E[i,j], { \beta_{EN}}  S_N[i,j]}
  \end{equation}
  for the traffic turning from east to north and
  \begin{equation}
    \label{eq:50}
    \begin{split}
      \Phi_N^{\text{in}}[i,j] & = \Phi_{NN}[i,j] + \Phi_{EN}[i,j] +
                                \Phi_{WN}[i,j] + \Phi_{SN}[i,j]\qquad
                                \forall\quad i,j\;, 
    \\
    \Phi_N^{\text{out}}[i,j] & = \Phi_{NN}[i,j] + \Phi_{NE}[i,j] +
                               \Phi_{NW}[i,j] + \Phi_{NS}[i,j] \qquad
                                \forall\quad i,j
    \end{split}
  \end{equation}
  for the first cardinal direction. The other directions are computed
  accordingly. Now the combined mixing flux is (for the north
  direction)
  \begin{equation}
    \label{eq:55}
    \Phi_N^\text{mix}[i,j] = \frac{1}{L[i,j]} \left(
      \Phi_N^{\text{in}}[i,j] - \Phi_N^{\text{out}}[i,j]\right)\;. 
  \end{equation}
\item As inflow and outflow are given for every minute, we also have
  to determine for the recent time step the actual minute before we
  retrieve the according values in each grid cell:
  \begin{equation}
    \label{eq:56}
    D_N^\text{source}[i,j],\dots, D_S^\text{source}[i,j],\qquad
    S_N^\text{sink}[i,j], \dots, S_S^\text{sink}[i,j] \qquad
    \forall\quad i,j\;. 
  \end{equation}
  With these and the demand and supply from equations~\eqref{eq:47}
  and~\eqref{eq:48} we get the desired inflow and outflow fluxes via
  equations~\eqref{eq:io27} and~\eqref{eq:io28} as
  \begin{equation}
    \label{eq:57}
    \Phi_N^\text{source}[i,j] =
    \min{D_N^\text{source}[i,j],\,S_N[i,j]},\quad
    \Phi^\text{sink}[i,j] =
    \min{D_N[i,j],\,S_N^\text{sink}[i,j]}
  \end{equation}
  for the northbound traffic. For the other cardinal directions, the
  formulas are likewise. The according combined io-flux reads as
  \begin{equation}
    \label{eq:58}
    \Phi_N^\text{io}[i,j] = \frac{1}{L[i,j]}
    \left(\Phi_N^\text{source}[i,j] -
      \Phi^\text{sink}[i,j]\right)\qquad \forall\quad i,j\;. 
  \end{equation}
\item The last step now is to sum up the advective, mixing, and
  inflow/outflow parts of the update, multiply it by the time
  step~\(\Delta t\) and add them to the old state:
  \begin{equation}
    \label{eq:60}
    \rho_N^\text{new}[i,j] = \rho_N^\text{old}[i,j] + \Delta t
    \left(\Phi_{N,\text{adv}}[i,j] + \Phi_N^\text{mix}[i,j]
      + \Phi_N^\text{io}[i,j] \right) \qquad \forall\quad i,j
  \end{equation}
  for the northbound fraction of the density. The other partial
  densities are treated in the same manner. 
\end{itemize}
In Section~\ref{sec:time-step-restr} we will discuss the choice of the
time step~\(\Delta t\) and a subcycling strategy for the inflow and
outflow terms. This means that we will modify the last step by
treating the io-flux separately using more intermediate steps if the
time step restriction for that part is much smaller than for the other
contributions, especially the advective part.

\section{Boundary treatment}
\label{sec:boundary-treatment}

In their paper, Tumash et~al.~\cite{TUMASH2022374} claim that no
boundary conditions need to be set due to the fact that incoming and
outgoing traffic can be modelled via sources and sinks at the
outermost intersections in the street network under
consideration. However in their published code~\cite{lyurlik_multidirectional_traffic_model}, they set homogeneous
Dirichlet conditions at the boundary. In this section, we discuss the
necessity of boundary values and a suitable way for setting them as
well as an advantageous choice for the size of  the computational
domain for a given street network.

\subsection{Differences between incoming and outgoing traffic}
\label{sec:diff-betw-incom}

The basic idea of Tumash et~al.~\cite{TUMASH2022374,tumash2021traffic}
is that vehicles can physically only enter or leave the network at
certain intersections. The network data provided with the code also
contain information whether an intersection has streets coming in from
outside of the domain of interest or vice versa. Thus, we know the
exact position of vehicles entering or leaving. Due to the usually
irregular shape of the street network, typically only four of these
intersections are located at the boundary, even when the rectangular
domain for computation is minimized. All other entry points are well
within the interior of the domain. Thus, at least at first glance, it
seems reasonable that inflow and outflow is modeled as point sources
(or sinks) at the location of the according intersection. However,
this only really holds true for incoming traffic.

Vehicles that enter the computational domain start out at the exact
position of the intersection.
A consequence of this is that, depending on the grid resolution, these \enquote{incoming} intersections may be over-emphasized in the numerical results, as seen in Figure~\ref{fig:resultate} on the extra fine grid, where vehicles are initially located in a grid cell that contains at least parts of streets. Beyond that, traffic seems to be more spread out, as vehicles spill over to cells that do not contain any parts of a street.
%
%
On the extra coarse grid, however, these
intersections are not pronounced at all. This is due to the fact that
every grid cell contains parts of several streets. Since we work with
averages over the cells, everything seems more leveled out.

With vehicles leaving the domain, the situation is different. Since
traffic during computation is not exactly bound to streets, there is
no guarantee that outgoing traffic exactly hits the cells with
\enquote{outgoing} intersections. On the one hand, this might improve with
coarser grids. On the other hand, coarser grids allow for larger
regions to be simulated. Once the region contains parts of the country
side, the roads and, thus, the intersections, are further apart from
each other. So, the issue reappears. Thus, we have to deal with
vehicles missing these \enquote{outgoing} intersections.

\subsection{Consequences for the boundary treatment}
\label{sec:cons-bound-treatm}

\begin{figure}
  \centering
    \includegraphics[width=.33\linewidth]{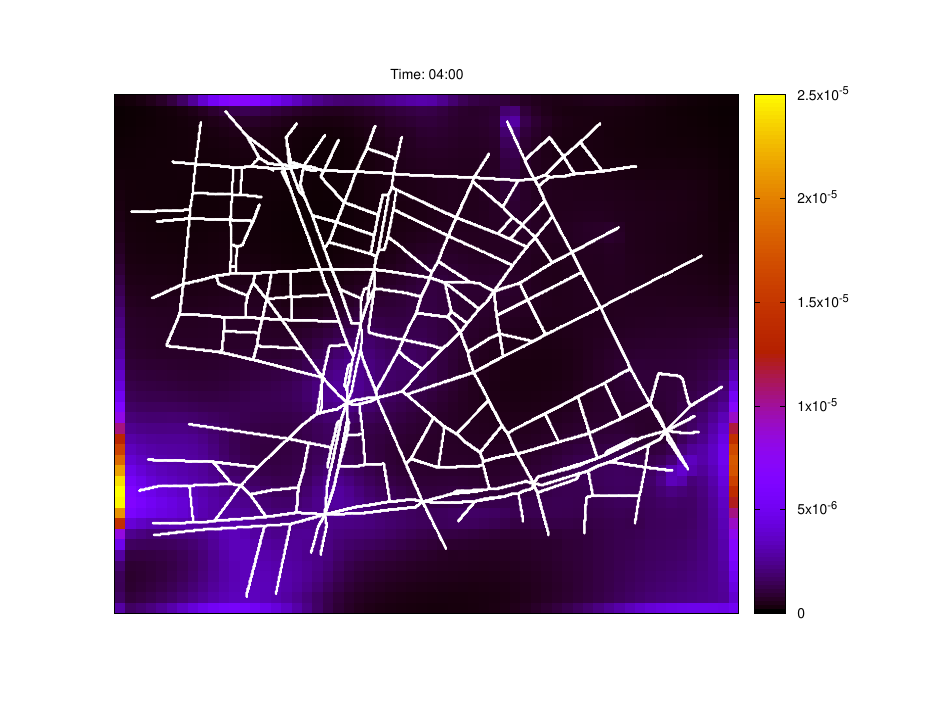}\hfill
    \includegraphics[width=.33\linewidth]{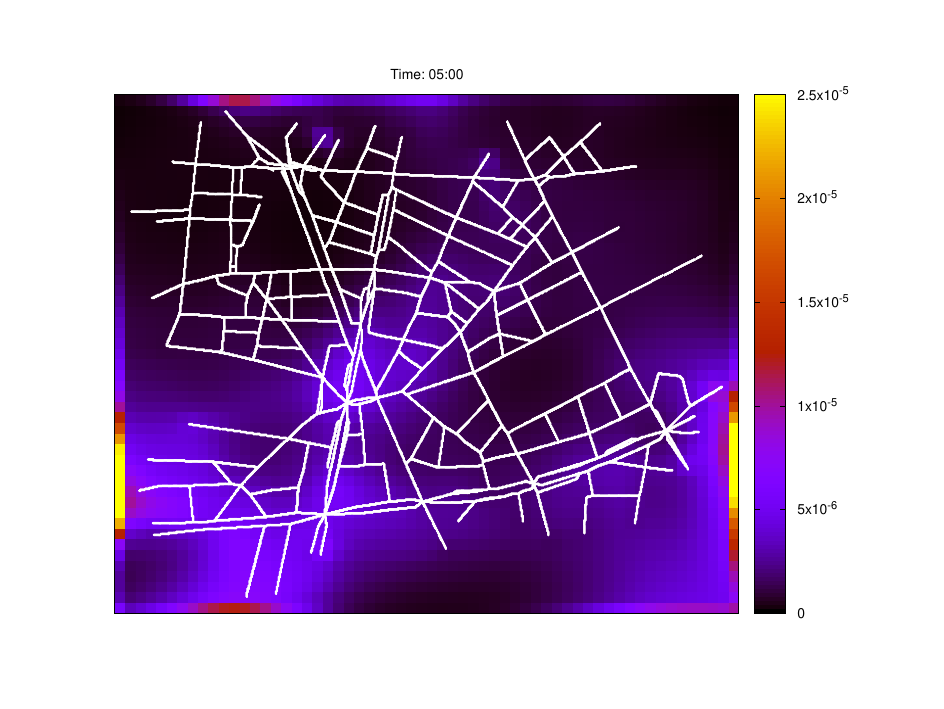}\hfill
    \includegraphics[width=.33\linewidth]{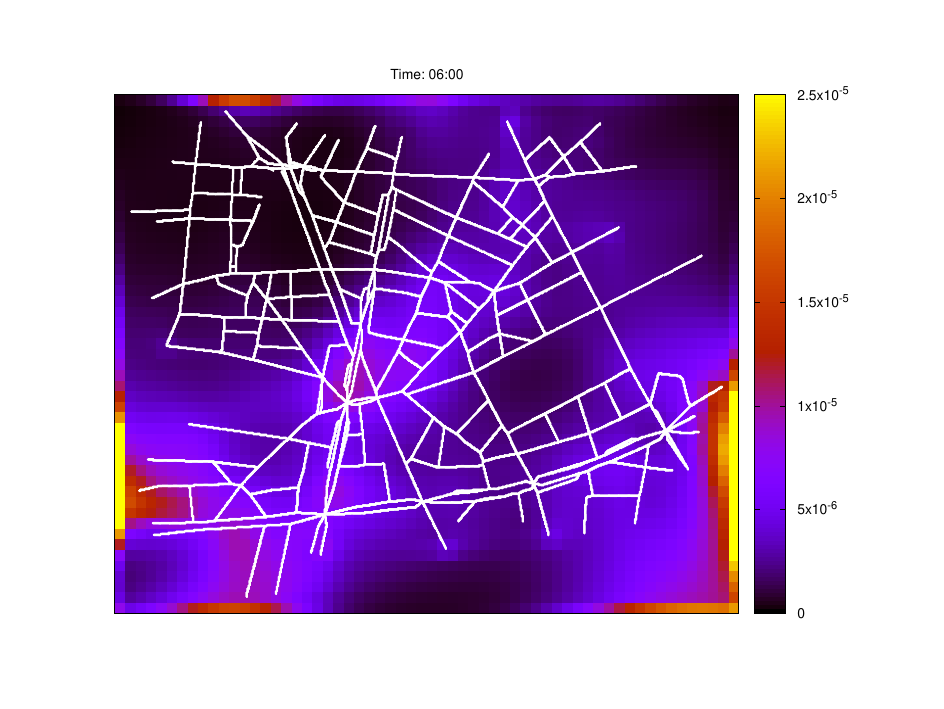}\\
    \includegraphics[width=.33\linewidth]{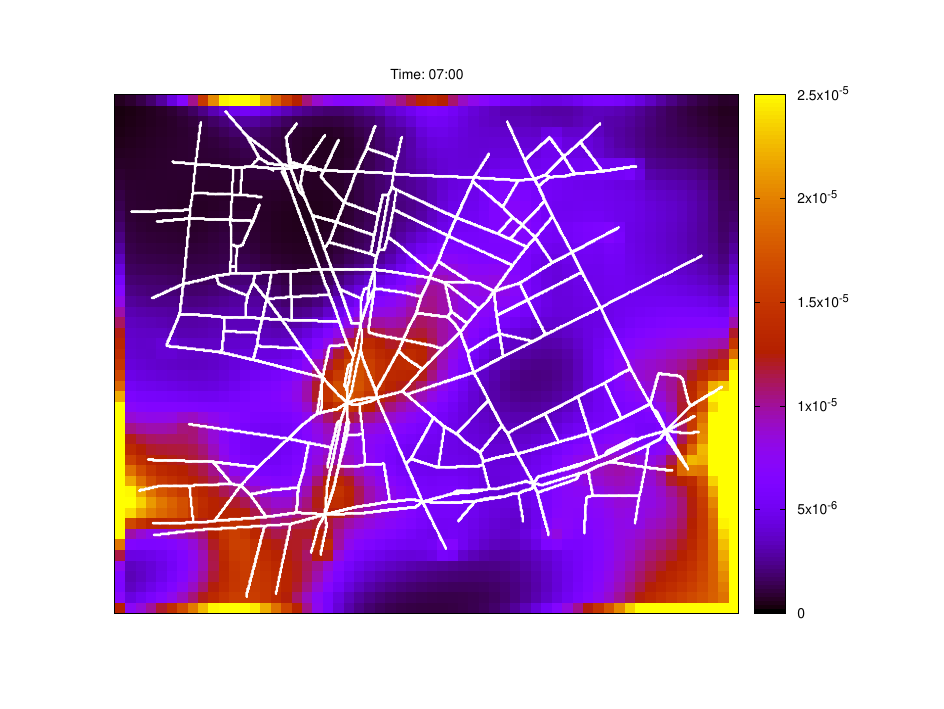}\hfill
    \includegraphics[width=.33\linewidth]{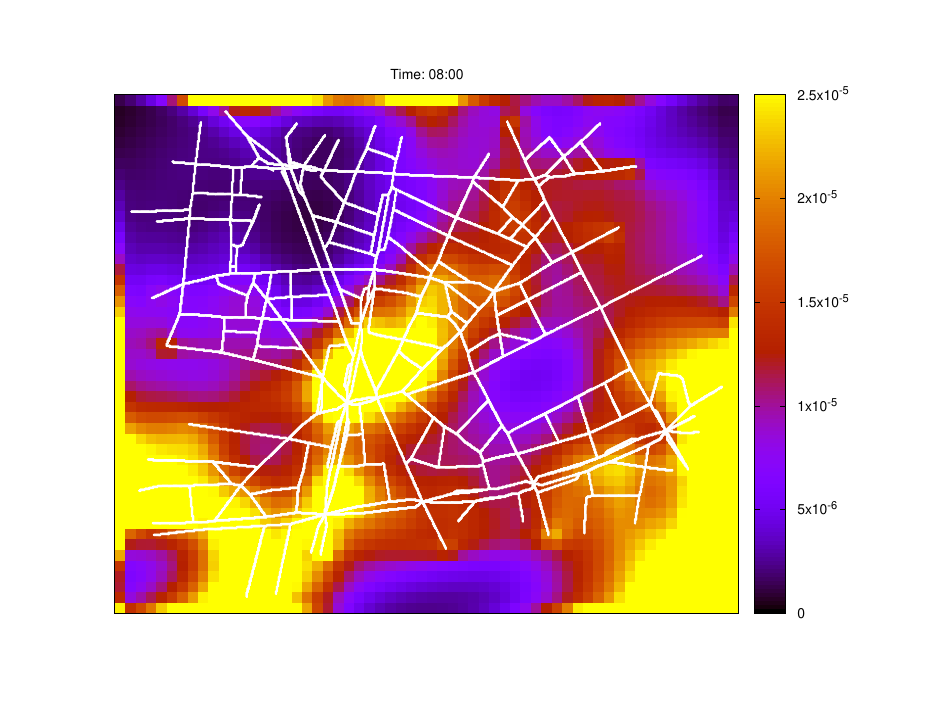}\hfill
    \includegraphics[width=.33\linewidth]{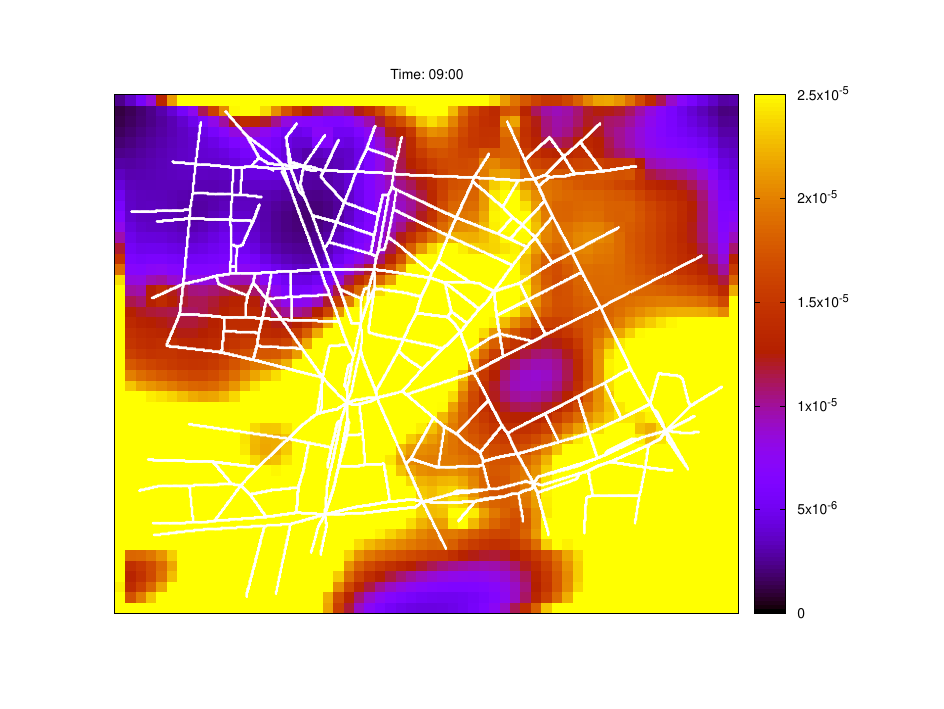}
    \caption[Solutions without transport over boundary]{Solution
      without transport across the boundary on a grid with
      61\(\times\)50 grid cells, hourly from 4\,am to 9\,am.}
  \label{fig:wrong-bc}
\end{figure}

Numerical experiments as shown in Figure~\ref{fig:wrong-bc} indicate
that it is necessary to implement boundary conditions that let
vehicles leave the computational domain without the need of exactly
hitting one of the \enquote{outgoing} intersections.
For explanation, when most 
%
%
outgoing vehicles miss the exact grid cells with the sink terms then,
as a consequence, they cannot leave the computational domain, and the
number of vehicles in the street network increases uncontrollably.
Thus, we follow the code by Tumash
et~al.~\cite{lyurlik_multidirectional_traffic_model} and implement
homogeneous Dirichlet conditions at the boundaries.

In order to avoid interaction of those vehicles missing the outgoing
intersections with the traffic in the street network, we implemented
another modification: Instead of choosing the smallest rectangle that
contains the street network, we enlarged the computational domain by a
few grid cells in every direction. We also decided not to smear the
sources for the inflow at all. For the intended simulations of large
areas with coarse grids, this would be counterproductive.
So, we just localized each
source exactly and solely in the containing grid cell.

The additional layers of grid cells allow for a simple implementation
of the desired homogeneous Dirichlet conditions: We simply set the
densities in the outermost layer to zero and do not update them during
the simulation. This corresponds with techniques from computational
fluid dynamics, where outside of the computational domain a layer of
so-called ghost cells is added. The results in
Figure~\ref{fig:resultate} are already computed with this approach.

\section{Time step restrictions and subcycling}
\label{sec:time-step-restr}

In order to compute time step restrictions, we have to consider
advection, mixing, and inflow/outflow separately. Any attempt to
compute it in a combined way would make things overly complicated. For
the advection part, we have to neglect the right-hand side of the
governing equations~\eqref{eq:1}, for the mixing, we ignore both the
source/sink fluxes and the left-hand side except for the time
derivatives. In the same way, for the sources and sinks, we only
consider the time derivatives together with the inflow/outflow fluxes.

\subsection{Time step size for advection}
\label{sec:time-step-size}

Since the averaged trigonometric terms have absolute values less or
equal one, we can neglect them in the computation of the time step
restriction. The wave speed then is given by the derivative of the
flux with respect to the density. The advective flux is given by
\begin{equation}
  \label{eq:13}
  \begin{split}
    \Phi(\rho) & = \min{D(\rho),S(\rho)}\\
               & = \min{v_\text{max}\rho,\abs{c_K}(\rho_\text{max} -
                 \rho), v_\text{max} \rho_c}\\[\medskipamount] 
               &= v_\text{max}\, \min{\rho, \frac{\gamma}{1-\gamma} 
                 (\rho_\text{max} - \rho), \gamma \rho_\text{max}}\;. 
  \end{split}
\end{equation}
With this, we find
\begin{equation}
  \label{eq:14}
  \max*[\rho]{\abs{\pd*{\Phi}{\rho}(\rho)}} = v_\text{max}\;. 
\end{equation}
Thus, for stability reasons, the (advective) CFL-number~\(c_\text{adv}\) has to satisfy
\begin{equation}
  \label{eq:15}
  c_\text{adv} = \frac{v_\text{max}\Delta t}{\Delta x} \leq 1
\end{equation}
for every direction in~\(\{N,E,W,S\}\).
This leads us to the overall CFL-condition
\begin{equation}
  \label{eq:16}
  \Delta t \leq \frac{\Delta x}{\max{\abs{v_{\text{max},N}},
      \abs{v_{\text{max},E}},\abs{v_{\text{max},W}},\abs{v_{\text{max},S}}}}\;,
\end{equation}
implying that we have to choose
\begin{equation}
  \label{eq:17}
  \Delta t = \frac{c_\text{adv}\cdot \Delta x}{\max{\abs{v_{\text{max},N}},
      \abs{v_{\text{max},E}},\abs{v_{\text{max},W}},\abs{v_{\text{max},S}}}}\qquad
  \text{with}\qquad 0<c_\text{adv}\leq 1\;. 
\end{equation}
For practical reasons, we will allow the code to decrease the time
step down to the next step size that fits with the output time steps,
making sure, all time steps are of the same size. Since above
calculation is done in one space dimension, we might have to further
lower the advective CFL-number~\(c_\text{adv}\) due to
multidimensional effects. Our numerical scheme resembles Donor Cell
Upwind (DCU) for which LeVeque~\cite{leveque} gives a proof of
stability for~\(c_\text{adv}\leq 1/2\). Based on heuristical considerations,
Toro~\cite{toro} gives a higher bound for which stability can still be
expected.

It is important to keep in mind that we did not build this condition
on the positivity of partial densities. But we can nevertheless expect
it to be satisfied in the absence of mixing or inflow and outflow
due to the use of the Godunov scheme, at least when applied in one
space direction. This can also be seen directly from the definition of
the flux function via demand and supply.

\subsection{Time step size for mixing}
\label{sec:time-step-size-1}

Although it is not necessary for the stability of the simulation, we
want to avoid both negative densities and densities exceeding the
maximal density. Therefore, we have to consider whether a time step
restriction for the mixing terms is necessary. If we want strict
positivity, we have to require it for all partial densities. Since the
actual density of cars in a grid cell is given by the sum of the
partial densities, we might accept the weaker requirement that the
summed up density in a cell is non-negative. If, as discussed above,
we neglect both advection and inflow/outflow, the equations reduce to
\begin{equation}
  \label{eq:18}
  \begin{split}
    {\rho_N}_t & = \frac{1}{L} \left( {\Phi_N^{\text{in}} -
               \Phi_N^{\text{out }}}\right)\\
  {\rho_E}_t & = \frac{1}{L} \left( {\Phi_E^{\text{in}} -
               \Phi_E^{\text{out }}}\right)\\
  {\rho_W}_t & = \frac{1}{L} \left( {\Phi_W^{\text{in}} -
               \Phi_W^{\text{out }}}\right)\\
  {\rho_S}_t & = \frac{1}{L} \left( {\Phi_S^{\text{in}} -
               \Phi_S^{\text{out }}}\right)\;.
  \end{split}
\end{equation}
If we sum these up and substitute equations~\eqref{eq:19}, everything
on the right-hand side cancels out, and we arrive at
\begin{equation}
  \label{eq:20}
  \pd{}{t}(\rho_N + \rho_E + \rho_W + \rho_S) = 0\;. 
\end{equation}
Thus, we expect no effect from the mixing on the positivity and
boundedness by the maximum of the combined density.

If we want to impose strict positivity, i.\,e.\ positivity for all
partial densities, we have to deal with the outgoing flux. For a
single direction, e.\,g.\ northwards, this leads to 
\begin{equation}
  \label{eq:21}
  {\rho_N}_t  = - \frac{1}{L}  \Phi_N^{\text{out }}
  =  - \frac{1}{L} \left( \Phi_{NN} + \Phi_{NE} + \Phi_{NW} + \Phi_{NS}\right)  \;.
\end{equation}
We now consider the case that all but one of the turning fluxes
vanish, e.\,g.\
\begin{equation}
  \label{eq:23}
  {\rho_N}_t  = - \frac{1}{L} \Phi_{NE} = \min{\alpha_{NE}D_N,\beta_{NE}S_E};. 
\end{equation}
If we assume the supply to be high enough, we get as worst case
\begin{equation}
  \label{eq:24}
  {\rho_N}_t  = - \frac{1}{L} \alpha_{NE} \rho_N v_\text{max}\;. 
\end{equation}
Since the only thing we know to be true for all turning ratios is the
boundedness by zero and one, we may as well drop it and, thus, receive
\begin{equation}
  \label{eq:25}
  {\rho_N}_t  = - \frac{1}{L} \rho_N v_\text{max}\;.
\end{equation}
Now, the condition for the stability of the explicit Euler method
applied to this resembles also the condition for positivity:
\begin{equation}
  \label{eq:26}
  \Delta t \leq \frac{L}{v_\text{max}}\;. 
\end{equation}
This condition has to be satisfied for each grid cell and for each
cardinal direction.  In our code, we use for simpler computation
\begin{equation}
  \label{eq:27}
  \Delta t \leq \underline L \cdot
  \underline{\left[\frac{1}{v_\text{max}}\right]}
  \qquad \left( \leq
    \underline{\left[\frac{L}{v_\text{max}}\right]}\right)\;, 
\end{equation}
where the underbar denotes the minimum over all grid cells, a notation
that we will also use in the next section for better readability.
By introducing a mixing CFL number, we arrive at 
\begin{equation}
  \label{eq:28}
  \Delta t = c_\text{mix} \cdot \underline L \cdot
  \underline{\left[\frac{1}{v_\text{max}}\right]} \qquad\text{with}
  \qquad 0 < c_\text{mix} \leq 1\;. 
\end{equation}

\subsection{Time step size for inflow and outflow}
\label{sec:time-step-size-2}

For an ODE 
\begin{equation}
  \label{eq:io1}
  \dot y =  \lambda y
\end{equation}
with a real parameter~\(\lambda\),
the stability condition for the explicit Euler method requires
\begin{equation}
  \label{eq:io2}
  -2 \leq \lambda \Delta t \leq 0\;.
\end{equation}

In the NEWS-model, we can---at least virtually---apply some kind
of operator splitting, which for the inflow and outflow terms results
in
\begin{equation}
  \label{eq:io3}
  \vec \rho_t = \frac{1}{L}\,\left[\vec \Phi^\text{source} (\vec \rho)
    - \vec \Phi^\text{sink} (\vec \rho)\right]\;.  
\end{equation}
Note that we dropped the index for the cardinal direction since the
equation looks the same for every partial density.  The fluxes for the
sources and sinks are given by equations~\eqref{eq:io27}
and~\eqref{eq:io28}.
Since all equations of this system are decoupled, and since there is
no space dependency, in every grid cell, we have to solve a scalar ODE
of the form
\begin{equation}
  \label{eq:io4}
  \rho_t =  \frac{1}{L}\,\left[\Phi^\text{source} (\rho)
    - \Phi^\text{sink} (\rho)\right]
\end{equation}
with
\begin{equation}
  \label{eq:io5}
  \Phi^\text{source} (\rho) = \min{ D^\text{source}, S
    (\rho)}\;,\qquad
  \Phi^\text{sink} (\rho) = \min{D (\rho), S^\text{sink}}  \;.
\end{equation}
Demand and supply are given by equations~\eqref{eq:io18}
and~\eqref{eq:io19}, their \(\rho\)-derivatives by~\eqref{eq:io20}
and~\eqref{eq:io21}.
In order to compute the stability condition for the time step, we
consider the linearized equation, whose flux, due to above
considerations, is defined piecewise. Since the cases~\(\rho<0\)
and~\(\rho > \rho_\text{max}\) can be excluded anyway, we are left
with two cases. The first one is~\(0\leq \rho \leq
\rho_\text{crit}\). Here, the equation can be written as
\begin{equation}
  \label{eq:io7}
  \rho_t =  \frac{1}{L}\,\left[\min{D^\text{source},v_\text{max}
      \rho_\text{crit}} - \min{v_\text{max} \rho, S^\text{sink}} \right]
\end{equation}
If no supply for outflow is given in the cell under consideration, the
contribution from the outflow flux vanishes, and we are, thus, left with a
constant. Also, if the supply is smaller than~\(v_\text{max} \rho\),
the same happens. Otherwise, we are left with
\begin{equation}
  \label{eq:io9}
  \rho_t =  \frac{1}{L}\,\left[\min{D^\text{source},v_\text{max}
      \rho_\text{crit}} - v_\text{max} \rho \right]\;,
\end{equation}
which is already linear. For the stability of the forward Euler
method, this requires
\begin{equation}
  \label{eq:io10}
  \frac{1}{L}\,v_\text{max}\,\Delta t \leq 2\qquad\Leftrightarrow\qquad
  \Delta t \leq \frac{2L}{v_\text{max}}\;. 
\end{equation}
The remaining case
is~\(\rho_\text{crit} < \rho \leq \rho_\text{max}\). For this, the ODE
becomes
\begin{equation}
  \label{eq:io11}
  \rho_t =  \frac{1}{L}\,\left[\min{D^\text{source},v_\text{max}\,
      \frac{\rho_\text{crit}(\rho_\text{max} - 
        \rho)} {\rho_\text{max} - \rho_\text{crit}}}
    - \min{v_\text{max}\,\rho_\text{crit}, S^\text{sink}} \right]
\end{equation}
If no demand to enter the street network exists in the according grid
cell, the inflow flux completely vanishes, and we are left with a
constant. Still, if the demand is low compared
to~\(v_\text{max}\, \frac{\rho_\text{crit}(\rho_\text{max} - \rho)}
{\rho_\text{max} - \rho_\text{crit}}\), we are also left with a
constant. Otherwise, the ODE becomes
\begin{equation}
  \label{eq:io12}
  \rho_t =  \frac{1}{L}\,\left[v_\text{max}\,
    \frac{\rho_\text{crit}(\rho_\text{max} - \rho)} 
    {\rho_\text{max} - \rho_\text{crit}} -
    \min{v_\text{max}\,\rho_\text{crit}, S^\text{sink}} \right]\;,
\end{equation}
the linearized version of which is
\begin{equation}
  \label{eq:io13}
  \rho_t =  \frac{1}{L}\,\left[\frac{-v_\text{max}\,\rho_\text{crit}}{\rho_\text{max}
      - \rho_\text{crit}}\,\rho - \min{v_\text{max}\,\rho_\text{crit},
      S^\text{sink}} \right]\;. 
\end{equation}
Thus, the stability condition becomes
\begin{equation}
  \label{eq:io14}
  \frac{1}{L}\,\frac{v_\text{max}\,\rho_\text{crit}}{\rho_\text{max}
      - \rho_\text{crit}}\,\Delta t \leq 2\qquad\Leftrightarrow\qquad
    \Delta t \leq  2L\cdot \frac{\rho_\text{max}
      - \rho_\text{crit}}{v_\text{max}\,\rho_\text{crit}}
    = \frac{2L}{v_\text{max}}\cdot \frac{\rho_\text{max}
      - \rho_\text{crit}}{\rho_\text{crit}}\;.
\end{equation}
If we now combine conditions~(\ref{eq:io10}) and~(\ref{eq:io14}), then
the stability condition becomes
\begin{equation}
  \label{eq:io15}
  \Delta t \leq  \frac{2L}{v_\text{max}}\cdot \min{1,\frac{\rho_\text{max}
      - \rho_\text{crit}}{\rho_\text{crit}}}\;. 
\end{equation}

Another question that is still open is the influence of the constants
on the validity of the resulting densities. Obviously the density has
to satisfy
\begin{equation}
  \label{eq:io16}
  0 \leq \rho \leq \rho_\text{max}\;. 
\end{equation}
To maintain this, it is necessary (but not always sufficient) that
\begin{equation}
  \label{eq:io17}
  \Phi^\text{source} (\rho) \Delta t \leq
  \rho_\text{max}\qquad\text{and}\qquad \Phi^\text{sink} (\rho) \Delta
  t \leq  \rho_\text{max}\;. 
\end{equation}
Since it is inconvenient to compute the resulting time step
restriction for every time step, and since we expect the state
dependent terms to be taken care of by the condition in
equation~(\ref{eq:io15}), we restrict ourselves to the terms that are
independent from the state. Thus, we have to deal
with~\(D^\text{source},\ S^\text{sink}\),
and~\(v_\text{max}\,\rho_\text{max}\). These lead to conditions
\begin{align}
  \frac 1 L \, D^\text{source}\,\Delta t & \leq
                                           \rho_\text{max}\;, \label{eq:io22}
  &\Leftrightarrow \qquad \Delta t & \leq \frac{L \rho_\text{max}}{D^\text{source}}\\
  \frac 1 L \, S^\text{sink}\,\Delta t & \leq
                                         \rho_\text{max}\;, \label{eq:io23}
  &\Leftrightarrow \qquad \Delta t & \leq \frac{L \rho_\text{max}}{S^\text{sink}}\\
  \frac 1 L \, v_\text{max}\,\rho_\text{max} & \leq
                                               \rho_\text{max}\;. \label{eq:io24}
                                               &\Leftrightarrow \qquad \Delta t & \leq \frac{L}{v_\text{max}}\;.
\end{align}
If we want the same time step for all grid cells, we have to use minima
and maxima for the parameters involved. If we denote minima and maxima
over all grid cells with underbars and overbars, we get for the
time step restriction induced by inflow and outflow
\begin{equation}
  \label{eq:io25}
  \Delta t \leq \underline L \min{\frac{2}{\overline{v_\text{max}}}\cdot \min{1,\frac{\underline{\rho_\text{max}}
        - \rho_\text{crit}}{\rho_\text{crit}}},
    \frac{ \underline{\rho_\text{max}}}{\overline{D^\text{source}}},
    \frac{ \underline{\rho_\text{max}}}{\overline{S^\text{sink}}},
    \frac{1}{\overline{v_\text{max}}}}\;.
\end{equation}
Since in the code,~\(\rho_\text{crit}\) is computed as a constant
multiple of the maximal density with factor~\(C\in (0,1)\), we can
rewrite this as
\begin{equation}
  \label{eq:io26a}
  \Delta t \leq \underline L \min{\frac{2}{\overline{v_\text{max}}}\cdot \min{1,\frac{1-C}{C}},
    \frac{ \underline{\rho_\text{max}}}{\overline{D^\text{source}}},
    \frac{ \underline{\rho_\text{max}}}{\overline{S^\text{sink}}},
    \frac{1}{\overline{v_\text{max}}}}\;.
\end{equation}
An improvement can be made by computing all four restriction terms for
each cell before taking their minimum over the grid. We also introduce
a small parameter~\(\varepsilon\) in order to avoid division by
zero. Furthermore, we employ some kind of CFL-number~\(c_\text{io}\leq
1\)
as a safety measure to address possible issues from nonlinearities. 
\begin{equation}
  \label{eq:io26}
  \Delta t \leq c_\text{io} \cdot \underline L
  \min{2\cdot\underline{\left[\frac{1}{v_\text{max}}\right]}\cdot
    \min{1,\frac{1-C}{C}},\, 
    \underline{\left[\frac{ \rho_\text{max}}{D^\text{source}+\varepsilon}\right]},\,
    \underline{\left[\frac{ \rho_\text{max}}{S^\text{sink}+\varepsilon}\right]},\,
    \underline{\left[\frac{1}{v_\text{max}}\right]}}\;.
\end{equation}
In order to be independent of time, we have to make sure
that~\(\frac{ \rho_\text{max}}{D^\text{source}}\)
and~\(\frac{ \rho_\text{max}}{S^\text{sink}}\) are not only minimized
over all grid cells, but also over all available input times.

\subsection{Subcycling for inflow and outflow}
\label{sec:subcycl-infl-outfl}

Now we have three different time step restrictions for the different
parts of the equations. The easiest way would be to take the minimum
of these and keep the numerics as in the original scheme by Tumash
et~al.~\cite{TUMASH2022374}. However, since the restrictions for
mixing and inflow/outflow don't depend on the grid resolution, for
coarse grids, this would prevent us from increasing the time step with
the grid spacing, resulting in an unnecessarily high computational
effort. With this respect, it would be optimal to treat all
three---advection, mixing, and inflow/outflow---separately in an
operator splitting, each with its own time step. But with regard to
equations~\eqref{eq:17},~\eqref{eq:28}, and~\eqref{eq:io26}, and
considering the fact that restriction~\eqref{eq:28} for the mixing
part is only necessary when we opt for strict positivity, i.\,e.\
positivity even of the partial densities, we decided that inflow and
outflow cause the most severe restriction for the time step on coarse
grids and, thus, restricted ourselves to a splitting in an advection
and mixing operator as well as an inflow/outflow operator. However, it is
still open for future work to do the full splitting.

The resulting time stepping strategy is as follows:
\begin{itemize}
\item If strict positivity is required, compute a time
  step~\(\Delta t_\text{general}\) using the minimum of the time steps
  resulting from equations~\eqref{eq:17} and \eqref{eq:28}.

  Otherwise, i.\,e.\ if positivity is only required for the summed up
  density, compute~\(\Delta t_\text{general}\) solely based on
  equation~\eqref{eq:17}.
\item If the resulting number of time steps per output cycle is no
  integer, reduce it so that we have equally sized time steps
  throughout the computation and exactly hit the output times. 
\item Compute the io time step~\(\Delta t_\text{io}\) according to
  equation~\eqref{eq:io26} and, if necessary, reduce it so
  that~\(\Delta t_\text{general}\) is an integer multiple of~\(\Delta
  t_\text{io}\), i.\,e.\
  \begin{equation}
    \label{eq:63}
    \Delta t_\text{general} = K\,\Delta t_\text{io}\qquad
    \text{with}\quad K \in \N. 
  \end{equation}
  In this way, also the inflow/outflow time steps are
  the same throughout the computation.
\item Then, first update the partial densities based on advection and
  mixing:
  \begin{equation}
    \label{eq:61}
    \rho_N^\text{intermediate}[i,j] = \rho_N^\text{old}[i,j] + \Delta t
    \left(\Phi_{N,\text{adv}}[i,j] + \Phi_N^\text{mix}[i,j] \right)
    \qquad  \forall\quad i,j
  \end{equation}
  Afterwards start the subcycling with~\(\rho_N^\text{\text{sub}\,0} =
  \rho_N^\text{old}\) and
  \begin{equation}
    \label{eq:62}
    \rho_N^{\text{sub}\,k}[i,j] = \rho_N^\text{\text{sub}\,k-1}[i,j] + \Delta t
    \Phi_N^\text{io}[i,j] \qquad \forall\quad i,j,\quad k=1,\dots,K\;.
  \end{equation}
\item Now we simply have to set
  \begin{equation}
    \label{eq:64}
    \rho_N^\text{new}[i,j] = \rho_N^{\text{sub}\,K}[i,j]\;.
  \end{equation}
\end{itemize}
Note, however, that this procedure takes the form of a
predictor-corrector method, implying that the in-cell demand and
supply have to be computed again for every subcycle. As a consequence,
when there is only one subcycle necessary, the computational effort is
higher than for the original unsplit scheme. This, in turn, implies
that the subcycling is only worthwhile when at least two subcycles are
needed, which mainly occurs on coarse grids.

\section{Numerical results}
\label{sec:numerical-results}

\subsection{The new code and the choice of parameters}
\label{sec:new-code}

In order to test our modifications, we set up a new code written in
Fortran. While the original code by Tumash
et~al.~\cite{lyurlik_multidirectional_traffic_model} relied on an
equal number of cells in both directions, we now allow for different
numbers of grid cells in the \(x\)- and \(y\)-direction. We deemed
this important to avoid distorted grid cells when the computational
domain is far from being a square. In fact, we chose the resolution in
\(x\)- and \(y\)-direction so that the grid spacing in both space
directions is about the same.
For the main loop, two variants are available: one with the original
unsplit algorithm as described in
Section~\ref{sec:numerical-treatment} and one with the subcycling
presented in Section~\ref{sec:subcycl-infl-outfl}. 
Naturally, the code also implements the time step calculation as
described in Section~\ref{sec:time-step-restr} in both variants, the
strict and the non-strict positivity condition. We can switch between
these variants via a logical parameter.

For all test cases, we set one minute (60\,s) as the upper bound for
the time step. Every 15~minutes, we output data. As mentioned before,
the time step size is adjusted in order to exactly meet the output
times. As a result, a slight change in the CFL-numbers might lead to a
major change of the time step size as can be seen in
Table~\ref{tab:strict-vs-nonstrict}.
%
%

The required parameters are chosen as in the original work by Tumash
et~al.~\cite{TUMASH2022374,lyurlik_multidirectional_traffic_model}:~At first~\(\mu = 0.02\) in
equation~\eqref{eq:39}.  The space needed per car in traffic jam
is~\(6\,m\), which leads to a maximal density of
\begin{equation}
  \label{eq:65}
  \rho_\text{max} = \frac{N}{6\,m}\;,
\end{equation}
where~\(N\) denotes the number of lanes. The ratio between critical
density and jam density in equation~\eqref{eq:8} is set
as~\(\gamma= 1/3\).
In addition, in some places we employed a small
parameter~\(\varepsilon = 10^{-8}\) to avoid division by zero.
Note that all parameters concerning the initial condition
in~\cite{TUMASH2022374} are not needed here, since we always start at
midnight with an empty street network. 

If not stated otherwise, the CFL-numbers are chosen
as~\(c_\text{adv}=0.5,\ c_\text{mix} = 0.57,\ c_\text{io} = 1.0\). The
choice for the mixing CFL-number is tuned so that in the comparisons
between strict and non-strict positivity in the next Section, there is
a notable difference in the time step size for advection and mixing
and, as a consequence, a different number of subcycling steps for
inflow/outflow. Whenever we use the original unsplit scheme, we
enforce all three conditions~\eqref{eq:17}, \eqref{eq:28},
and~\eqref{eq:io26}. In this case, the---compared
to~\(c_\text{io}\)---relatively low mixing CFL-number may require a
smaller time step than the non-strict positivity computation, even
when there is only one io-subcycle.

\subsection{Strict vs non-strict positivity condition for mixing}
\label{sec:strict-vs-non}

\begin{figure}
  \centering
  \includegraphics[width=.48\linewidth]{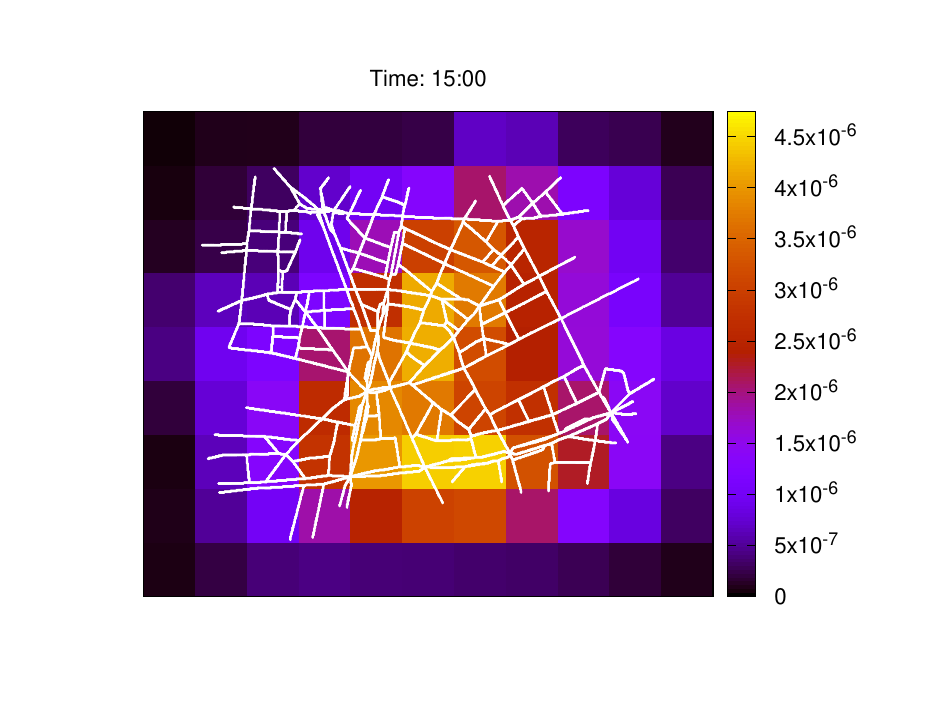}\hfill
  \includegraphics[width=.48\linewidth]{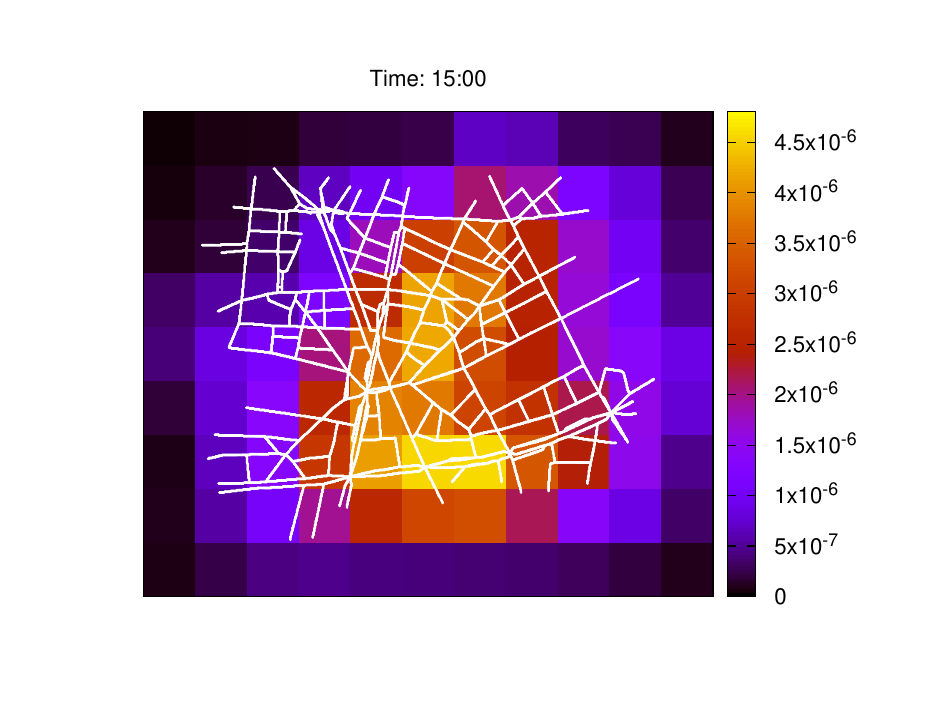}\\
  \includegraphics[width=.48\linewidth]{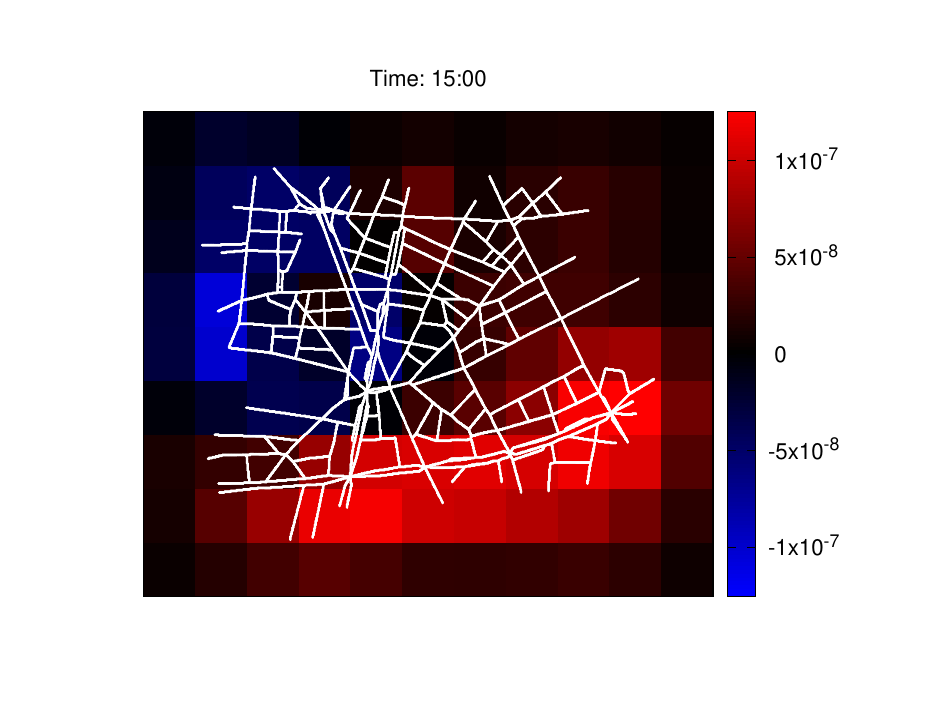}
  \caption[Non-strict vs.\ strict positivity]{Numerical results at
     3\,pm on a 12\(\times\)10 grid with advective
    CFL-number 0.5. Top left:  non-strict positivity; top right: strict
    positivity with mixing CFL-number 0.57; bottom row: difference between
    strict and non-strict positivity.}
  \label{fig:strict-vs-nonstrict}
\end{figure}

Since the subcycling in general only kicks in on a coarse grid, we
show a comparison for 12\(\times\)10 grid cells in
Table~\ref{tab:strict-vs-nonstrict}. The according numerical results
for the latter two together with the difference between the strict and
the non-strict version are shown in
Figure~\ref{fig:strict-vs-nonstrict}. It becomes obvious that
requiring strong positivity leads to a lower general time step, thus,
increasing the computational effort. Although the number of io-cycles
per time step is also decreased, the complete number of io-cycles is
still slightly higher than without the requirement for strict
positivity of the partial densities. On the other hand, the resolution of
the numerical solution is slightly improved. The magnitude of the
difference between both solutions is roughly a factor 50 lower than
the numerical solution itself.

\begin{table}
  \centering
  \begin{tabular}{SSSScc}
    \multicolumn{1}{c}{\(c_\text{adv}\)} & \multicolumn{1}{c}{\(c_\text{mix}\)} &
                                                      \multicolumn{1}{c}{\(c_\text{io}\)}
  & \multicolumn{1}{c}{\(\Delta t\)} & \multicolumn{1}{c}{subcycles
                                       for io} &
                                                                       \multicolumn{1}{c}{steps
                                                                       per
                                                                       output}
  \\ \midrule
  0.45 & \multicolumn{1}{c}{---} & 1.0 & 9.8901 & 2 & 91 \\
  0.5  & \multicolumn{1}{c}{---} & 1.0 & 14.0625 & 3 & 64 \\
  0.5 & 0.57 & 1.0 & 7.9646 & 2 & 113\\ \bottomrule
  \end{tabular}
  \caption[With and without strict positivity.]{CFL-numbers and
    resulting time steps for 12\(\times\)10 grid cells with and
    without strict positivity. Actual number for \(c_\text{mix}\)
    indicates strict positivity.}
  \label{tab:strict-vs-nonstrict}
\end{table}


Note that we implemented a check for violations of positivity and
boundedness by~\(\rho_\text{max}\). In the case of strict positivity,
i.\,e.\ the case where we enforce positivity for partial densities by
setting a mixing CFL-number, we check all partial densities,
otherwise, we only check the summed up density in each grid cell. In
the case of the bounds being violated, we consider the computation as
failed, and the code stops.



\subsection{Effect of subcycling on results}
\label{sec:effect-subcycl-resul}

In this Section, we want to discuss the effects of the subcycling
without considering the issue of strict vs.\ non-strict
positivity.
Because the largest general time step is obtained without strict positivity of all partial densities, we employ this case for the computations with subcycling.


We want to compare this version with the version where the general
time step is lowered down to the minimum of the mixing and the io time
step, without any subcycling, going back to the original unsplit
scheme.
We want to compare this version with the one where the general time step is lowered down to the minimum of the mixing and the io time step, without any subcycling, thus, going back to the original unsplit
scheme.

As mentioned above, for the unsplit scheme, we employ also the
strict positivity condition.


\begin{figure}
  \centering
  \includegraphics[width=.33\linewidth]{frame_nonstrict05_0060}\hfill
  \includegraphics[width=.33\linewidth]{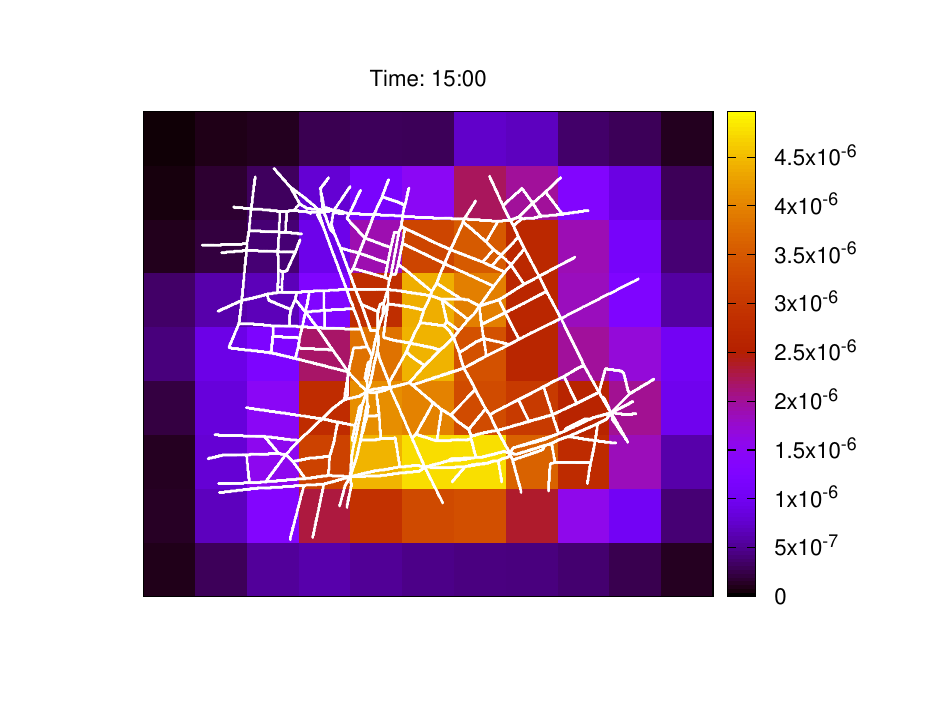}\hfill
  \includegraphics[width=.33\linewidth]{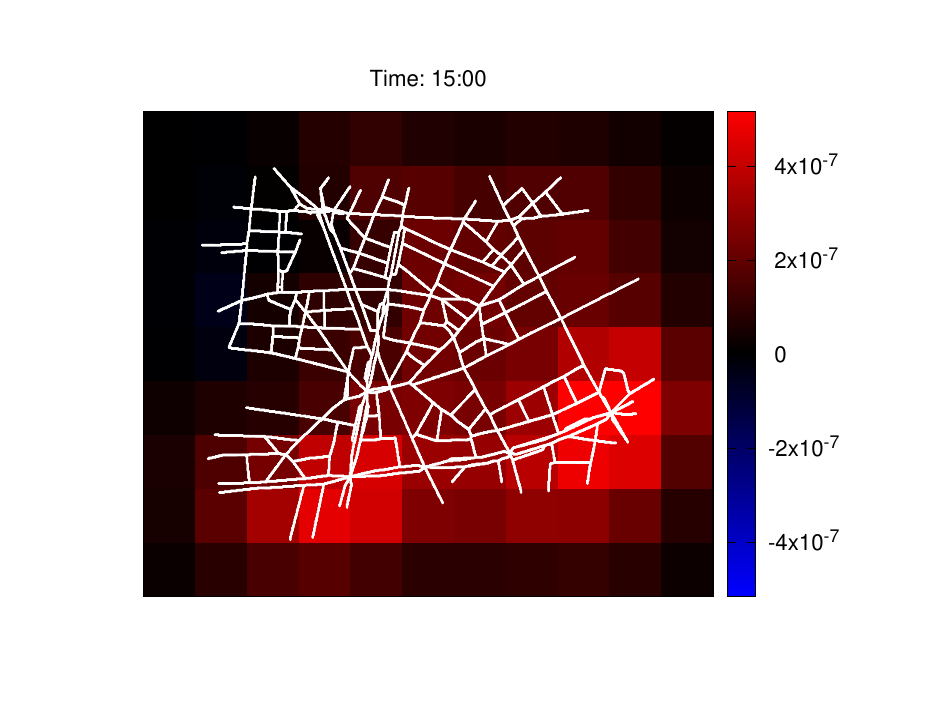}\\
  \includegraphics[width=.33\linewidth]{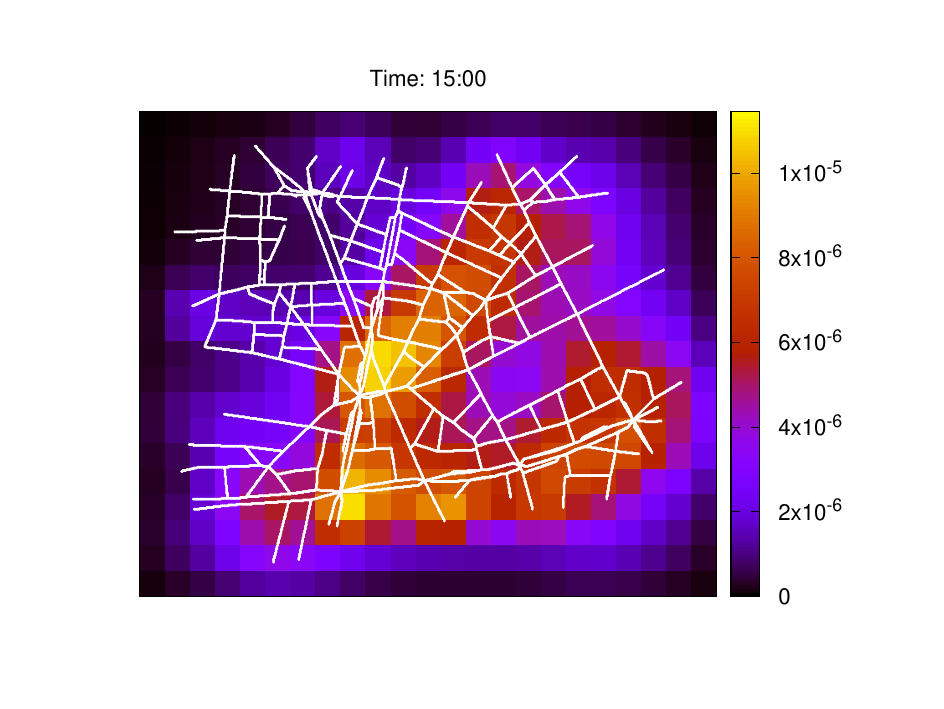}\hfill
  \includegraphics[width=.33\linewidth]{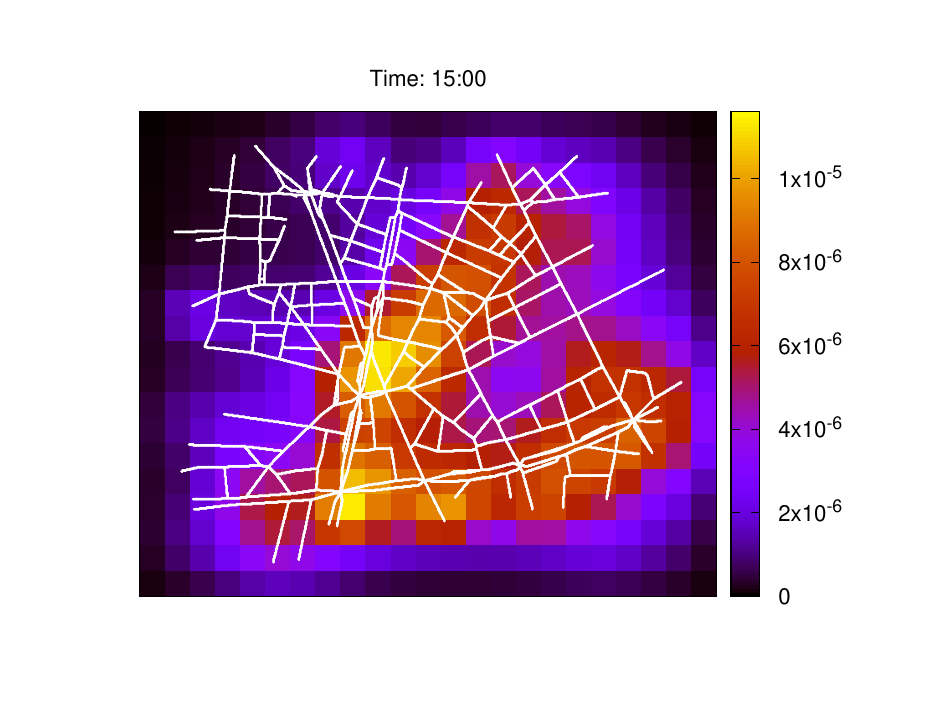}\hfill
  \includegraphics[width=.33\linewidth]{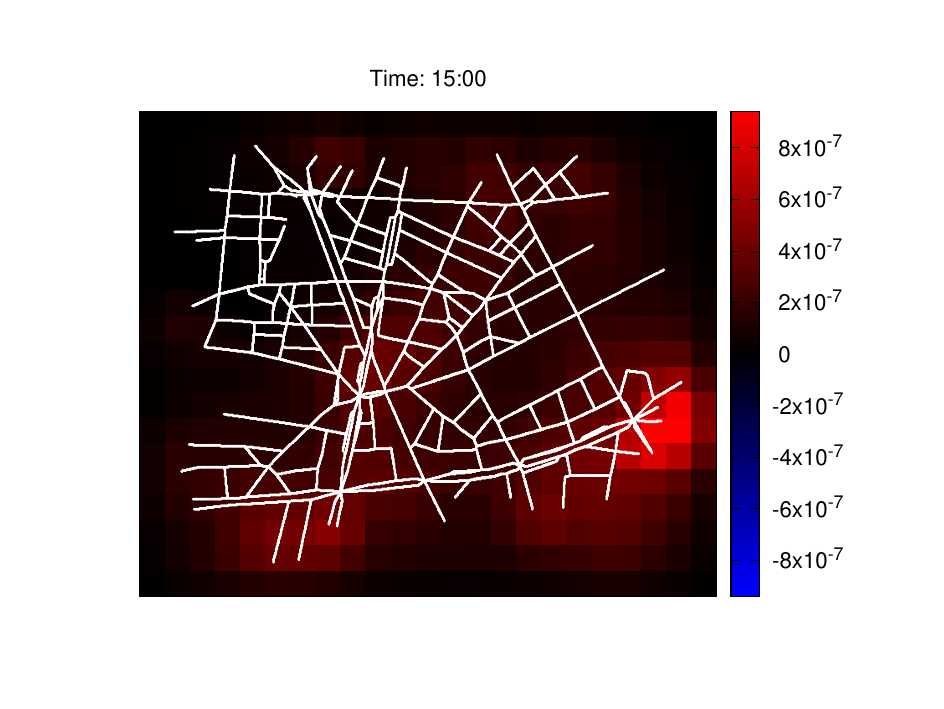}\\
  \includegraphics[width=.33\linewidth]{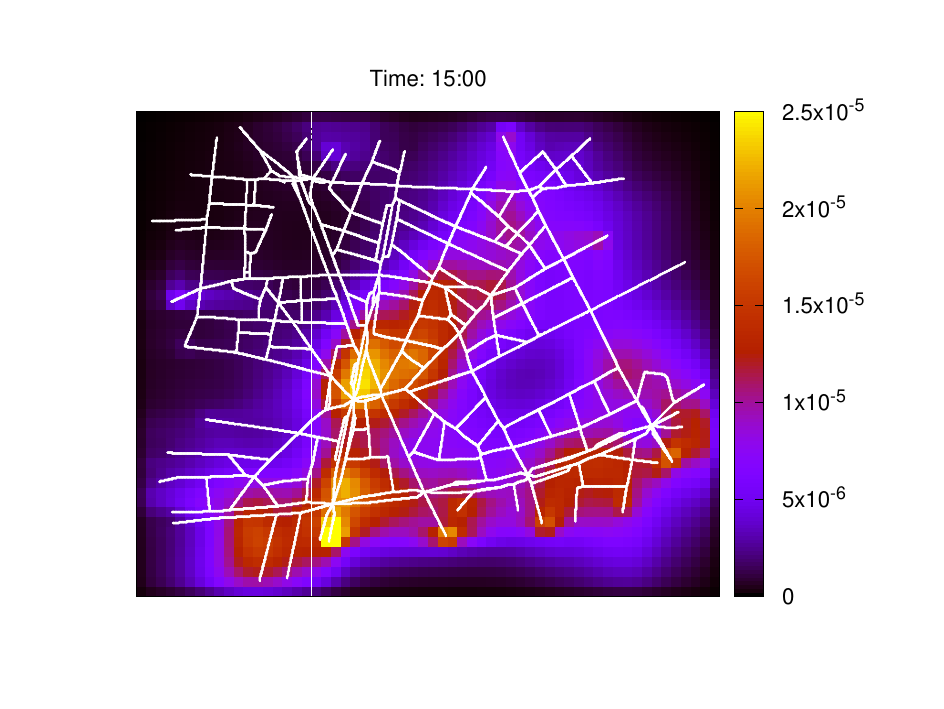}\hfill
  \includegraphics[width=.33\linewidth]{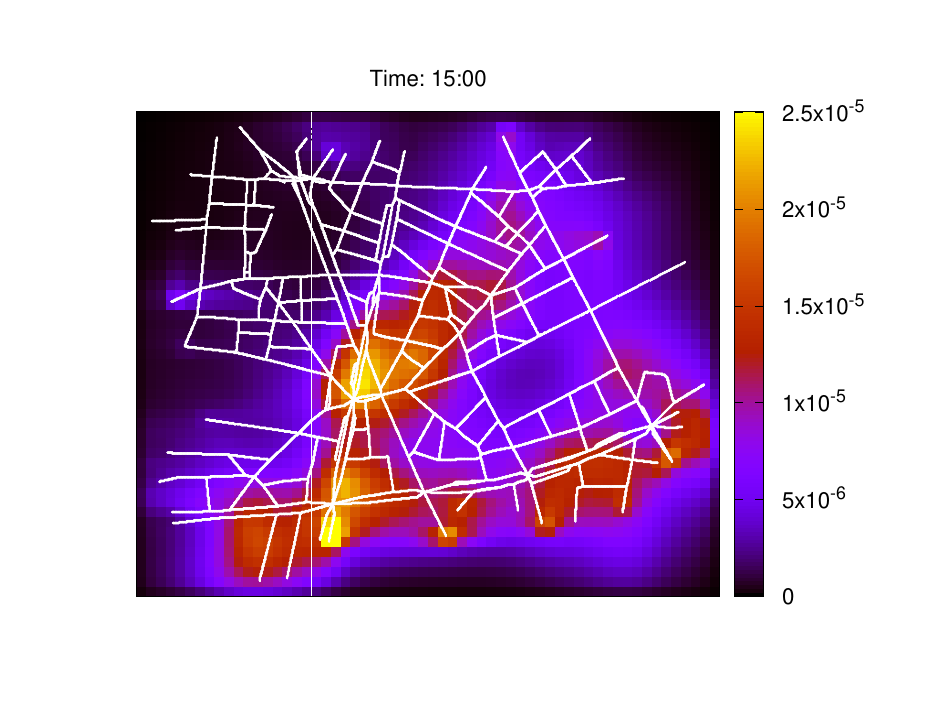}\hfill
  \includegraphics[width=.33\linewidth]{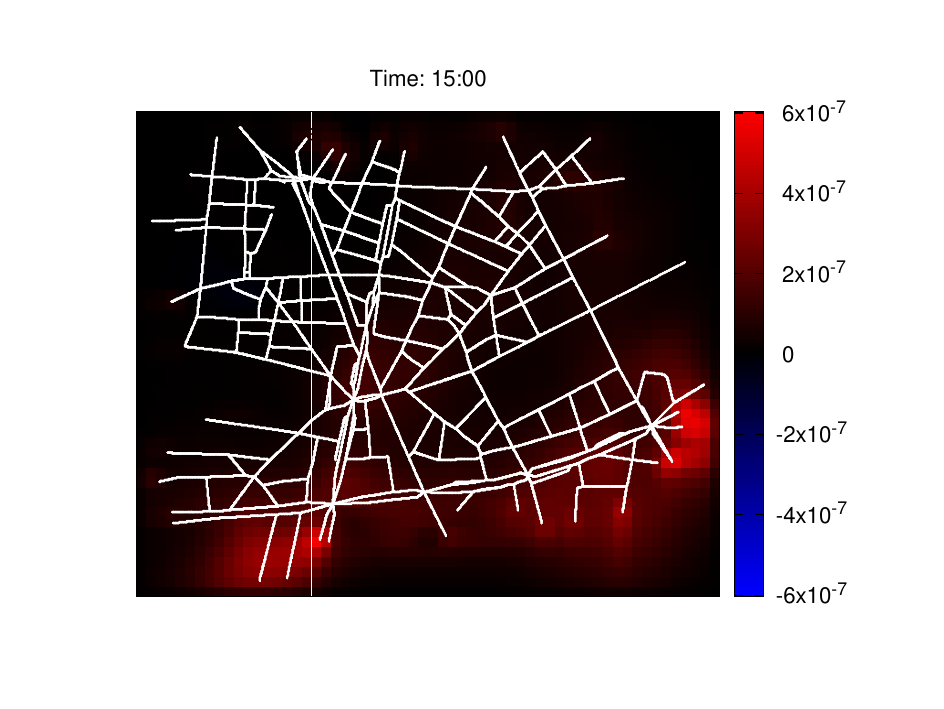}\\
  \caption[split vs.\ unsplit]{Numerical results at 3\,pm on a
    12\(\times\)10 grid, a 24\(\times\)20 grid, and a 61\(\times\)50
    grid with advective CFL-number 0.5 (from top to bottom). From
    left to right: split with non-strict positivity, unsplit,
    difference between unsplit and split version.}
  \label{fig:split-unsplit-mult}
\end{figure}

Figure~\ref{fig:split-unsplit-mult} shows results for three different
grid resolutions. The 12\(\times\)10 grid was already used in the last
section. Here, we used three subcycles in the operator splitting
version of the method. On the 24\(\times\)20 grid, this number reduced
to two. On the finer grids we tested, only one io-substep per time
step was performed. Thus, the computational effort even increased
compared to the unsplit scheme. For the 24\(\times\)20 grid, however,
there was no significant difference in computation time between the
split and unsplit method.
The results for the two variants look almost indistinguishable at
first sight. However, difference plots (right column) show some
distinctive features, especially near the intersections with high
inflow and outflow. But the effect on the interior parts of the
computational domain is minimal. Thus, we conclude that, in terms of
precision, the difference between the results of both variants is
negligible, wile the decision, whether to use the split or unsplit
scheme, has to be made based on the computational cost. On fine grids,
there is no gain from the operator splitting, while on coarse grids,
the computation time can be reduced via subcycling for inflow and
outflow. This is in accordance with our goals: To save computation
time on coarse grids as required for fast simulations of large
geographical regions.


\section{Conclusion and outlook}
\label{sec:conclusion-outlook}

In this study, we showed how the implementation of the 2d continuous
traffic model by Tumash
et~al.~\cite{TUMASH2022374,lyurlik_multidirectional_traffic_model} can
be improved, especially in terms of the computational effort. We
introduced an a priori time step calculation that allows for
considerably larger time steps, especially on coarse grids. Even for a
610\(\times\)500 grid and with positivity for all partial densities
required, we were able use a slightly larger time step than is
suggested by Tumash et~al.\ for any grid resolution, in particular the
60\(\times\)60 grid they used for their numerical tests. Furthermore,
we introduced a subcycling for the inflow and outflow that lead to a
considerable decrease in computational effort on coarse grids. The
deviations between the results of the different versions are way
smaller than the modeling error itself. Besides these improvements, we
adjusted the boundary treatment on the basis of the code provided by
Tumash et~al.~\cite{lyurlik_multidirectional_traffic_model}.  Choosing
a larger computational domain, allowed for vehicles to leave the
domain even if they missed the cells with \enquote{outgoing}
intersections.
At last, using 2d-densities of the original 1d traffic densities
improved comparison between results with different grid spacing.

There are still some open questions with the 2d continuous model: We
relied on empirical data for the turning ratios provided by Tumash
et~al.~\cite{lyurlik_multidirectional_traffic_model} bundled with
their code. These might be a good basis as long as the sole purpose of
the model is to control traffic in a restricted part of an urban
street network. In this case, the purpose would justify the effort for
collecting the required data. But for larger street networks not only
the effort to collect the data is extremely high, also the data 
themselves are less useful if dealing with traffic planning and
possible changes to street networks. Thus, it would be desirable to
find a way to directly derive estimates for the turning ratios from
the properties of the streets and intersections
themselves. Furthermore, there are many more modes of transport than
motorized vehicles. Thus, we not only need additional equations for
additional modes, but also a model for the interaction between the
modes and a way to treat mode changes. Finally, also sources and sinks
for people starting and ending their travel within the computational
domain have to be implemented. Thus, we can conclude that, while the
2d continuous model is a promising starting point for further research
and development, there remains some work to be done before the model
can be fully used in traffic planning.

\subsection*{Acknowledgements:}
\label{sec:acknowledgements:}

I would like to thank Dr.~Michael Landes for his invaluable assistance
in proofreading the manuscript. His meticulous review greatly improved
the readability and clarity of the manuscript.



\subsection*{Funding:}
\label{sec:funding:}

This work was supported by the German Federal Ministry of Education and
Research (BMBF)
[grant number 05M22ICA].




\bibliographystyle{amsplain} \bibliography{2dmod-paper}

\end{document}